\documentclass[a4paper,11pt]{article}
\pdfoutput=1 

\usepackage{jcappub} 

\usepackage[T1]{fontenc} 

\usepackage{epstopdf}
\usepackage{hyperref, graphicx, slashed, xcolor, amsmath}
\usepackage{natbib}
\usepackage{tensor}
\allowdisplaybreaks

\newcommand{\MeV}{\text{MeV}}

\abstract{A strongly self-interacting component of asymmetric dark matter can collapse and form compact objects, provided there is an efficient mechanism of  energy evacuation. If the dark matter quantum number is not completely conserved but it is slightly violated due to some new physics e.g. at the Planck scale, dark matter particles can annihilate into Standard Model particles. Even tiny annihilation cross sections are sufficient to create observable luminosities. We demonstrate that these dark matter annihilations can trigger radial pulsations, causing a characteristic time modulation of the luminosities produced. We argue that in this scenario, the spectral features along with the properties of the oscillation can create a unique discovery signal for such objects in the sky.}

\title{Dark Matter Annihilation from Pulsating Dark Stars}
\author[a]{Chris Kouvaris}
\author[a]{Dimitris Zavitsanos}
\affiliation[a]{Physics Division, National Technical University of Athens, 15780 Zografou Campus, Athens, Greece}  
\begin{document}
\maketitle
\flushbottom
\section{Introduction}
Although the Collisionless Cold Dark Matter paradigm (CCDM) aligns with  large-scale structure observations, several small-scale structure issues create tension between CCDM and observations. One of the aforementioned issues is the core-cusp problem in galaxies~\cite{Moore:1994yx,Burkert:1995yz,deBlok:2001rgg}. Although N-body numerical simulations of  CCDM particles produce a cuspy dark matter (DM) density profile, observational data from rotational curves in dwarf galaxies that are dominated by DM indicate  a rather flat core. Another related issue is the so-called diversity problem~\cite{Oman:2015xda,Zentner:2022xux}. Although several galaxies exhibit the same asymptotic rotational velocities far from their centers, their velocities can differ substantially in their corresponding inner cores. Furthermore there is the ``too big to fail problem"~\cite{Boylan-Kolchin:2011lmk,Papastergis:2016}. Once again, N-body numerical simulations predict massive subhalos for our galaxy that are too big not to be able to have star formation and therefore being visible to us. All of the above issues can have different resolutions. For example N-body simulations might not be so accurate if we do not consider properly the back reaction of baryonic matter to the system. However, a very compelling explanation of all the aforementioned problems of CCDM can be the existence of DM self-interactions. For example it is easy to see how this works for example in the core-cusp problem: a  DM cuspy profile can be flattened due to extensive scattering among DM particles, thus being compatible with the observation of flat cores in dwarf galaxies.  Yet another reason pointing towards DM self-interactions is the existence of supermassive black holes up to $10^9 M_{\odot}$ ($M_{\odot}$ being a solar mass) usually located at the center of galaxies. Based on  current understanding of accretion rates into black holes and stellar formation timescales, it seems that starting from an astrophysical black hole of tens solar masses after the collapse of a supermassive star, there is  not enough time for the black holes to grow to the current observed masses. DM self-interactions can alleviate this problem too. A strongly self-interacting minor component of DM can collapse forming either black holes or massive compact objects early on that act as seeds that through merger or accretion can grow to $10^9 M_{\odot}$~\cite{Pollack:2014rja}. 

In this paper, we assume a strongly self-interacting component of DM  to study potential signals that can present themselves as compact DM objects. We  assume that this self-interacting DM component is asymmetric in nature. Asymmetric DM is an alternative to the Weakly Interacting Massive Particle (WIMP) paradigm. In the WIMP scenario, DM abundance is determined by a competition between DM annihilations and the expansion of the Universe. In the case of asymmetric DM, there is a conserved quantum number associated with DM playing the role of baryon number in hadrons~\cite{Nussinov:1985xr,Barr:1990ca,Gudnason:2006yj,Ryttov:2008xe,Kaplan:2009ag,March-Russell:2011ang,Buckley:2011ye,Davoudiasl:2011fj,Graesser:2011wi,Bell:2011tn} (for a review see~\cite{Petraki:2013wwa}). At some stage of the evolution, a mechanism (that can be similar to baryogenesis) creates an asymmetry between the population of DM particles and antiparticles. DM annihilations deplete the population of the antiparticles and therefore the Universe is left with the component in excess which cannot be further annihilated since the stability of the DM particle is protected by the conserved quantum number.  In this paper we will implement both elements mentioned above. We will consider a strongly asymmetric self-interacting component of DM. Self-interactions are essential because they provide the mechanism of evacuating energy from the system and facilitate the collapse and formation of the compact dark objects which we will call from now on as dark stars. However, on top of that we will allow small violations of the DM baryon-like number via annihilations of the DM particles to Standard Model (SM) particles and in particular in our considered scenario to photons. Although these  DM annihilations will be so small to inflict stability issues for dark stars or to cause observable signals from DM annihilations in the halo, the large compactness of dark stars with densities that can be potentially even larger than those of a neutron star, enhance dramatically the rate of annihilation that takes place inside the dark star. 

Strongly self-interacting asymmetric DM can collapse, eventually forming black holes or dark stars if there is an efficient way of evacuating energy from the system, thus facilitating the collapse. One possible way for achieving that is via DM self-interactions where for example two DM particles can scatter off resulting to a high energetic particle that escapes from the ensemble, thus lowering the overall energy of the remaining system which through thermalization and virialization will contract. Another possibility is the evacuation of energy via the emission of dark Bremsstrahlung radiation~\cite{Chang:2018bgx}. In this case DM couples to a dark photon and by emitting the latter, energy is lost and the system contracts. The whole process has been proven to lead to fragmentation of the collapsing self-gravitating clump
and it ends only once: i) the hoop conjecture is satisfied and a black hole is formed, or ii) the DM particles become degenerate and develop Fermi pressure (if DM is in the form of fermions), or iii) the mean free path of dark photons becomes smaller than the size of the clump which means that dark photons do not escape without rescattering and thus  returning back to the system part of their energy, or iv) DM-DM repulsions provide enough pressure to halt the collapse. Since the nature of DM is unknown (whether for example DM particles are fermions or bosons), there is a vast parameter space regarding DM and dark photon masses, and strength coupling between DM and dark photon. As a result, a very wide range of objects can be produced ranging from massive black holes to different types and masses of dark stars. The possibility of asymmetric  DM forming  dark stars was studied first in~\cite{Kouvaris:2015rea} for the fermionic DM and in~\cite{Eby:2015hsq} for bosonic. The possibility of forming admixed DM-baryon stars has also been studied~\cite{Tolos:2015qra,Deliyergiyev:2019vti,Mukhopadhyay:2015xhs,Brito:2015yfh, Mukhopadhyay:2016dsg,Cardoso:2019rvt,Maselli:2019ubs,Kain:2021hpk,Jimenez:2021nmr,Dengler:2021qcq,Ryan:2022hku,Collier:2022cpr,Sen:2022pfr,Cassing:2022tnn,Diedrichs:2023trk, Emma:2022xjs,Gartlein:2023vif,Giangrandi:2024qdb}. Furthermore DM can accumulate onto neutron stars and form cores at the center of the latter affecting potentially the equation of state (EoS) and stability of the stars~\cite{Kouvaris:2010vv,deLavallaz:2010wp,Kouvaris:2010jy,Kouvaris:2011fi,Kouvaris:2012dz,Kouvaris:2013kra,Bramante:2013hn, Bertoni:2013bsa,Bramante:2014zca,Bramante:2017ulk, Baryakhtar:2017dbj,Kouvaris:2018wnh,Garani:2018kkd,McKeen:2018xwc,Nelson:2018xtr,Garani:2019fpa,Ivanytskyi:2019wxd,Garani:2020wge,Karkevandi:2021ygv,RafieiKarkevandi:2021hcc,Giangrandi:2022wht} (see also~\cite{Bramante:2023djs} for a comprehensive review).

Several aspects of dark stars have been already explored. Dark stars can produce gravitational waves~\cite{Maselli:2017vfi}, electromagnetic radiation via either kinetic mixing~\cite{Maselli:2019ubs} or they can begin to radiate, producing outbursts of high luminosities~\cite{Kamenetskaia:2022lbf} due to accretion of interstellar gas of protons and electrons. 

In this paper we will consider perturbations from the equilibrium configurations of dark stars, focusing in radial oscillations. We study how the potential emission of electromagnetic radiation is affected by the pulsation, and we discuss the prospects of detection. 
The paper is organized as follows: In sec. \ref{2} we review the potential structure, Equation of State (EoS) and the profile of dark stars. In sec. \ref{3} we study the radial oscillations of these objects in the framework of general relativity. We study DM annihilations and we derive their related spectrum in sec. \ref{4}. In sec. \ref{5} we show how luminosity varies as a function of time if the dark star oscillates. In sec.~\ref{6} we study the conditions under which DM annihilations trigger 
the radial oscillations via the so-called $\epsilon$-mechanism. We conclude in sec. \ref{7}.

\section{Structure of Dark Stars}\label{2}
We focus on the scenario where DM particles are fermions that exhibit a self-interaction (repulsive or attractive) mediated by either a scalar or a gauge boson (dark photon). We assume the mediator has a mass and therefore induces a Yukawa type of potential between two DM particles of the form
\begin{equation}
V_{Y}(r)=\pm \frac{g^{2}}{4\pi}\: \frac{e^{-m_{\phi} r}}{r},
\end{equation}
where $g$ is the coupling strength between DM and mediator, $m_\phi$ is the mediator mass, and $r$ is the distance between the two DM particles. If the characteristic size of the star (i.e. its radius) is much larger than the Compton wavelength of the mediator $R>>m_\phi^{-1}$, one can easily calculate the total energy density of the Yukawa interaction inside the star which is~\cite{Shapiro:1983du} $  \rho_Y=\pm \;\frac{n^{2}}{2z^{2}}$, where $n$ is the DM number density inside the dark star and $z=m_\phi/g$. The contribution of the Yukawa interaction to the pressure is given via
\begin{equation}
p_{Y}=n^2\frac{d}{dn}\left ( \frac{\rho_{Y}}{n} \right ).
\end{equation}
Therefore in our setup the dark star settles into a configuration where gravity is balanced by the Yukawa pressure (if interactions are repulsive) and the Fermi pressure due to the degeneracy of DM. Throughout we will assume that the temperature of the dark star is much smaller to the DM chemical potential (which is ultimately related to the DM density) and therefore we will ignore the temperature dependence, effectively setting $T=0$. This is similar to what happens in neutron stars and it can be a posteriori justified by estimating the chemical potential of the DM particles. Based on the above discussion, the EoS of the dark star is parametrically given by
\begin{align}
\label{rho}
    \rho(x)&=\frac{m_{\chi}^{4}}{8\pi^{2}}\left[x\sqrt{1+x^{2}}(2x^{2}+1)-\ln{ \left(  x+\sqrt{1+x^{2}} \right)}\right] \pm \frac{n^{2}(x)}{2z^{2}}\\
  \label{pres}
    p(x)&=\frac{m_{\chi}^{4}}{8\pi^{2}} \left[x\sqrt{1+x^{2}}\left(\frac{2}{3}x^{2}-1\right)+\ln{ \left(  x+\sqrt{1+x^{2}} \right)}\right] \pm \;\frac{n^{2}(x)}{2z^{2}},
\end{align}
where $x=p_F/m_{\chi}$ with $p_F$ and $m_{\chi}$ being the Fermi momentum and mass of the DM particle respectively. The first term in both equations above correspond to the energy and pressure contribution of the kinetic energy of the degenerate Fermi gas~\cite{Shapiro:1983du}, while the second term arises as mentioned from the contribution of the Yukawa interaction. Since we consider a degenerate Fermi gas, the number density of DM $n$ is related to the corresponding Fermi momentum via
\begin{equation}
  n=\frac{m_{\chi}^{3}}{3 \pi^{2}} x^{3},
  \label{numberd}
\end{equation}
where we have assumed that the DM particle has spin $1/2$ and therefore 2 spin polarizations. Throughout the paper we use natural units i.e., $\hbar=c=1$.

We solve the  Tolman-Oppenheimer-Volkoff (TOV) equation, with the above EoS, to get the density profile, mass and radius of the dark star. However since we will study perturbations from  equilibrium configurations, it is instructive to review the basic steps for the derivation of the  TOV equation, which describes the hydrodynamic stability of the star in the framework of general relativity. To derive the TOV equation we need to solve Einstein's equations  $R\indices{_\mu_\nu}-\frac{R}{2}g\indices{_\mu_\nu}=8\pi G T\indices{_\mu_\nu}$, assuming that the energy-momentum tensor is that of an ideal fluid
$T_\nu^\mu=\rm diag [\rho, -p,-p,-p]$. The most general static and spherically symmetric metric is
\begin{equation}
ds^{2}=-e^{2\nu(r)}dt^{2}+e^{2\lambda(r)}dr^{2}+r^{2}d\theta^{2}+r^{2}\sin^{2}{\theta}d\phi^{2}.
\label{eq:metric}
\end{equation}
Substituting the above into the Einstein equations and using the conservation of energy-momentum ($\nabla_\mu T^{\mu \nu}=0$), we obtain
\begin{align}
\lambda'&= \frac{1-e^{2\lambda}}{2r}+4\pi G re^{2\lambda}\rho\\
\nu'&= \frac{e^{2\lambda}-1}{2r}+4\pi Gre^{2\lambda}p\\
p'&= -(\rho+p)\nu',
\end{align}
where primed quantities are derivatives with respect to $r$. The metric outside  the star must be the Schwarzschild one. To this end, we define
\begin{equation}
e^{2 \lambda(r)}={\left( 1-\frac{2G m(r)}{r} \right)}^{-1},
\end{equation}
where $m(R)=M$ can be identified as the star's total mass. We must also satisfy the following boundary condition
\begin{equation}
e^{-2\lambda(R)}=e^{2\nu(R)} \rightarrow \nu(R)=-\lambda(R).
\end{equation}
This enforces the compatibility with the Schwarzschild metric and will allow us to determine the value of the $\nu$ metric function.
Changing from the variable $\lambda$ to $m$ we get the system of equations describing the star's equilibrium
\begin{align}
  \frac{dm}{dr}&=4\pi \rho r^{2},\nonumber\\
  \frac{dp}{dr}&=-\frac{G m\rho}{r^2}\left (1+\frac{p}{\rho} \right )\left (1+\frac{4\pi r^3 p}{m} \right ) \left (1-\frac{2Gm}{r}\right )^{-1},\nonumber\\
\frac{d\nu}{dr}&=G\frac{4\pi r^{3}p+m}{r(r-2Gm)}.
\label{system}
\end{align}
The first equation corresponds to the conservation of mass in the nonrelativistic limit. The second equation is the widely known TOV equation. The first factor in the right hand side (i.e. $-Gm\rho/r^2$) is
the Newtonian result for hydrostatic equilibrium inside the star. All the factors in the parenthesis are corrections introduced by general relativity.
The system \eqref{system} together with EoS \eqref{rho}, \eqref{pres} can be solved in order to obtain all the desired functions $m(r)$, $\rho (r)$, $\nu (r)$ and $p(r)$.  This system of differential equations is accompanied with the following initial conditions:
\begin{enumerate}
  \item $m(0)=0$, stating simply that the mass enclosed at $r=0$ is zero.
  \item $\nu(0)=-1$, this value is chosen arbitrarily. Only the derivative of $\nu$ enters the equations, so we have the freedom to choose some initial value to start the integration. To compute $\nu$, after solving the system we shift the values of $\nu$ so that the boundary condition $\nu(R)=-\lambda(R)$ is satisfied.
        \item $\rho(0)=\rho_{c}$, each value of the central energy density $\rho_c$ gives a different star.
        \item $p(0)=p_{c}$, this is fixed from the EoS once we have chosen a value for $\rho_c$.
\end{enumerate}
As a sideline technical remark, integrating out \eqref{system} is problematic if we start the process exactly from the center of the star, due to the $r=0$ terms in the denominators of the expressions.
 An easy way to deal with this problem is to start the integration from a sufficiently small $r_0/R\approx 10^{-9}$, rather than at exactly $r=0$. To a very good approximation we assume that the conditions at $r=0$ and $r_0$ are the same. The initial conditions at $r_0$ are easily obtained from these at $r=0$ by Taylor expansion to second order
\begin{align}
m(r_0) &=  \frac{4}{3}\pi \rho_c r_0^{3}\nonumber\\
p(r_0) &=  p_c - 2\pi(\rho_c+p_c)(p_c+\frac{1}{3}\rho_c)r_0^{2}\nonumber\\
\nu(r_0) &= \nu_{c} + 2\pi(p_c+\frac{1}{3}\rho_c)r_0^{2},
\end{align}
where all quantities with index $c$ refer to values at $r=0$.
We should terminate the integration when the pressure reaches exactly zero. However  in practice we terminate numerically the integration once the pressure takes values below a tiny ($\sim 10^{-10}$) fraction of the core pressure.
At each step of the integration we need $\rho (r)$ at that specific radius, which we obtain from the EoS and the value of $p(r)$ calculated through the integration scheme. However we do not know directly $\rho$ as a function of $p$ or vice versa. Instead we know both $p$ and $\rho$ as functions of $x$. In practice we
 calculate both quantities for various $x$ and then using an interpolation (cubic spline), we construct the functions $p(\rho)$ and  $\rho(p)$.

\begin{figure}[htbp!]
  \centering
  \includegraphics[scale=1.0]{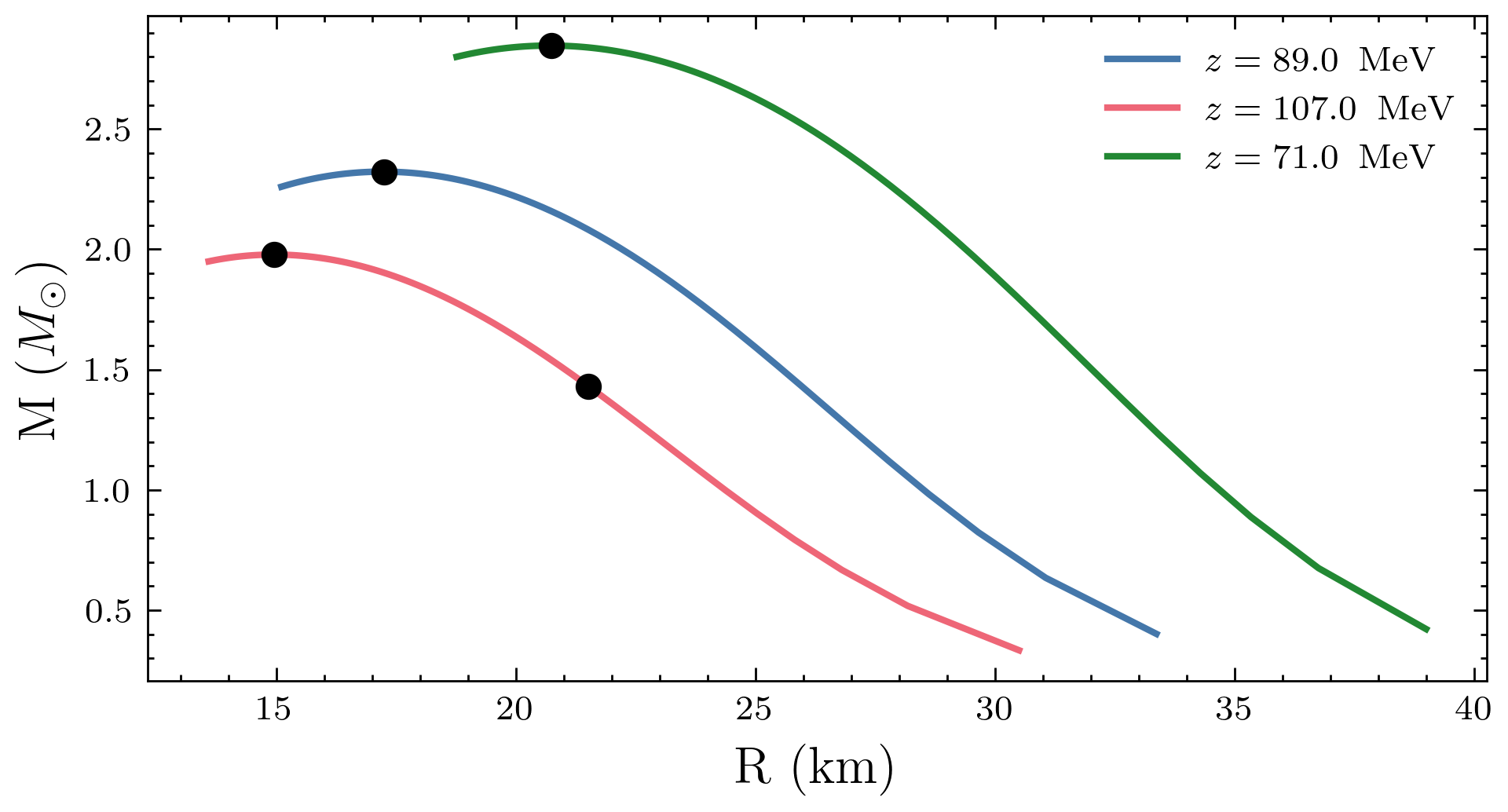}
  \caption{Mass-Radius relations of stars with fermionic DM of mass ($m_{\chi}=1$ GeV) and three different values of $z$ that parametrize the Yukawa interactions. The points at the top of each curve denote the star's configuration of maximum mass. All curve points to the right of these maximum mass configurations correspond  also to stable dark stars. The second point on the red curve corresponds to the fourth studied configuration of Table~\ref{tab:gaussian_fit}.\label{fig:MR}}
\end{figure}

In~Fig.\ref{fig:MR} we present the mass-radius relation for three families of stars characterized by the variables $z$ and $m_{\chi}$, which fully describe the EoS. For each pair of $z$ and $m_{\chi}$, there is a whole family of stars with different central densities and overall mass. As one can see, smaller masses correspond to larger radii. There is a maximum mass for each case denoted by a dotted point. This mass is similar in nature to the Chandrasekhar mass limit in white dwarfs. The solution points to the left of the maximum mass are unstable configurations. This is anticipated in analogy with white dwarfs and neutron stars but we will also verify it by investigating the frequency $\omega$ of the radial oscillation modes. These unstable solutions have $\omega^2<0$, meaning the modes grow exponentially rather than oscillating.
For the analysis of oscillations, we must have all the quantities describing the equilibrium. Therefore, for each star, we store the functions $m(r),\rho(r),\lambda(r),\nu(r)$. Regarding $\nu(r)$, we recall that we need to subtract an appropriate constant from the calculated result to obtain the correct value $\nu(R)$ which should match that of the Schwarzschild exterior solution at $r=R$. This constant is given by $\nu_{R}=\nu_{\text{calculated}}(R)-\frac{1}{2} \ln \left( 1-\frac{2G M}{R}\right)$, and thus, $\nu(r)=\nu_{\text{calculated}}(r)-\nu_{R}$.
Additionally, we need the adiabatic index \(\Gamma_1=\frac{\rho+p}{p} \frac{dp}{d\rho}\). The derivative \(\frac{dp}{d\rho}\) is calculated using the interpolation we performed for $p(\rho)$. Note that $\Gamma_{1}$ is not necessarily constant throughout the star in general and therefore it is a function of $r$.

\section{Radial Oscillations}\label{3}
We proceed with the study of radial oscillations around equilibrium configurations that satisfy the TOV equation. This is interesting for several reasons. Firstly, radial oscillations can be excited by  internal processes such as the $\epsilon$- and $\kappa$-mechanisms in ordinary stars. In particular as we will argue in sec. \ref{5}, DM annihilations even in small amounts can naturally induce radial oscillations via  the $\epsilon$-mechanism. Additionally, even in the absence of an intrinsic mechanism for generation of radial oscillations, a hyperbolic encounter of a dark star with another massive object (e.g. a black hole) can easily excite non-radial oscillations which through various dissipative processes induced by shear viscosity or heat conductivity can in the non-linear regime cause also a nonzero amplitude for radial oscillations. Given that non-radial perturbations generally lead to emission of gravitational waves, while radial oscillations do not, it is possible that radial oscillations might be the only modes surviving for long periods. Obviously in the absence of a mechanism sustaining  the pulsation, the radial oscillations will eventually die out. However, the absence of gravitational wave emission, can prolong the lifetime of these oscillations and their chance to be detected.

 Radial oscillations have been firstly studied in the context of Newtonian variable stars by Eddington, with a generalization in the context of general relativity by Chandrasekhar. We follow the formalism of \cite{Misner1973}.
 In order to study radial oscillations, we need to perturb the metric~\eqref{eq:metric} and the energy-momentum tensor around the equilibrium values. The perturbed quantities are
\begin{align}
\nu(t,r)&=\nu_{0}(r)+\delta \nu(t,r)\\
\lambda(t,r)&=\lambda_{0}(r)+\delta \lambda(t,r)\\
\rho(t,r)&=\rho_{0}(r)+\delta \rho(t,r)\\
n(t,r)&=n_{0}(r)+\delta n(t,r)\\
p(t,r)&=p_{0}(r)+\delta p(t,r),
\end{align}
where $n$ is the DM number density. We denote equilibrium quantities with the index $0$.  We describe the oscillation of a fluid element located initially at $r$ in the (unperturbed) equilibrium configuration, with a radial displacement from $r$ by $\xi(r,t)$.
We  assume that the time dependence of the above perturbations is harmonic in nature
\[\xi(r,t)=\xi(r)e^{-i\omega t}.\]
The functions $\delta \nu$, $\delta \lambda$, $\delta \rho$, $\delta n$, and $\delta p$ introduced above, describe the variation of quantities  measured by an observer fixed at a point $r$; these are referred to as \textit{Eulerian}. There is an equivalent description known as \textit{Lagrangian}, which describes the system from the view of an observer following the  motion of an individual fluid element. We will denote Lagrangian variations with  capital $\Delta$. One can easily switch  from one description to the other, using the following
\begin{align*}
  \Delta\nu(t,r)&=\nu(t,r+\xi)-\nu_{0}(t,r)=\nu(t,r)+\nu'(t,r)\xi -\nu_{0}(t,r)\\
  &=\delta \nu(t,r)+(\delta\nu'+\nu'_{0})\xi\approx \delta \nu(t,r)+\nu'_{0}\xi.
\end{align*}
Similar equations hold also for the other quantities i.e., $\lambda$, $\rho$, $n$ and $p$ which can easily be deduced from the above equation if we replace $\nu$ by any of the quantities mentioned.
The extra term arises because, in the Lagrangian description, the observer moves with the fluid element. In the general case, this term is given by the Lie derivative along the flow defined by $\xi$, because we have to bring the quantities at the same point to compare them.
We assume that the perturbation is small enough so that we can retain only the linear terms of the variations.

The conservation of DM number gives \cite{Misner1973}
\begin{equation}\label{eq:lagden}
\Delta n=-n_{0} \left[ r^{-2}e^{-\lambda_{0}} \left( r^{2}e^{\lambda_{0}}\xi \right)'+\delta \lambda \right].
\end{equation}
For a dark star that does not possess DM annihilations, the radial oscillations are adiabatic in nature. The existence of DM annihilations can make the oscillation non-adiabatic, although as we will argue in sec. \ref{5}, it is quasi-adiabatic. The adiabatic index of the system is
\[\Gamma_1=\frac{n}{p} \frac{\Delta p}{\Delta n}.\]
Solving the above for $\Delta p$ and using \eqref{eq:lagden} we get
\begin{equation}\label{eq:lagpress}
\Delta p=\Gamma_1 \frac{p_{0}}{n_{0}}\Delta n=-\Gamma_1 p_{0} \left[ r^{-2}e^{-\lambda_{0}} \left( r^{2}e^{\lambda_{0}}\xi \right)'+\delta \lambda \right].
\end{equation}
Given the above discussion regarding the quasi-adiabatic oscillation, energy conservation through the first law of thermodynamics (ignoring the small heat exchange) gives
\begin{equation}\label{eq:lagen}
\Delta \rho= \frac{\rho_{0}+p_0}{n_0}\Delta n=-(\rho_{0}+ p_{0}) \left[ r^{-2}e^{-\lambda_{0}} \left( r^{2}e^{\lambda_{0}}\xi \right)'+\delta \lambda \right].
\end{equation}
Perturbing Einstein's equations and in particular the ${rt}$ and ${rr}$ components, we get respectively
\begin{align}
\delta\lambda&=- \left( \lambda'_{0}+\nu'_0 \right)\xi=-4\pi (\rho_{0}+p_0)r e^{2\lambda_{0}}\xi\\[2ex]
    \delta\nu'&=- 4\pi \Gamma_1 p_0 r^{-1}e^{2\lambda_{0}+\nu_{0}} \left( r^{2} e^{-\nu_{0}}\xi \right)'+\left[p'_{0}r-(\rho_{0}+p_{0}) \right]4\pi e^{2\lambda_{0}}\xi.
\end{align}
We have a set of equations that give $\delta \nu'$, $\delta \lambda$, $\delta \rho$ and $\delta p$ in terms of $\xi$ and the unperturbed quantities. The relativistic Euler equation supplemented with the above set of equations can be recast in the form~\cite{Misner1973}
\begin{equation}
W\ddot{\zeta}=(P\zeta')'+Q\zeta, 
\label{main_eq}
\end{equation}
where we define
\begin{align}
  \zeta& \equiv r^2e^{-\nu_0} \xi\nonumber\\
  W& \equiv \left( \rho_0+p_0 \right)r^{-2}e^{3\lambda_0+\nu_0}\nonumber\\
  P& \equiv \Gamma_1 p_0r^{-2}e^{\lambda_0+3\nu_0}\nonumber\\
  Q&\equiv e^{\lambda_0+3\nu_0}\! \left[\frac{(p'_{0})^2}{\rho_0+p_0}r^{-2}\! -4p'_{0}r^{-3} -8\pi \left( \rho_{0}+p_0 \right)p_0r^{-2}e^{2\lambda_0} \right].
  \label{P}
\end{align}
Taking into account the assumed harmonic time dependence of $\xi$, Eq.~\eqref{main_eq} transforms to
\begin{equation}\label{eq:master}
  (P\zeta')'+(Q + \omega^{2} W) \zeta=0.
\end{equation}
Recall that $\omega$ is the angular frequency of the time dependence of $\xi$.
This equation is accompanied by two boundary conditions, one at the center and one at the star's radius. At the center, $\delta \rho$ and $\delta p$ must be finite. This leads to the condition that $\xi/r$ must be finite or zero as $r \rightarrow 0$.
The second boundary condition is related to the fact that the pressure outside the star must be zero; therefore, we impose
\begin{equation}
   \Delta p= -\Gamma_1 p_{0}r^{-2}e^{\nu_{0}} \zeta' \rightarrow 0~ \text{as}~r \rightarrow R.
    \end{equation}
    In practice, this translates to $p_{0}(R)\zeta'(R)=0$. 
  Recall that $R$ is the radius of the star.
Equation~\eqref{eq:master} along with the initial condition for $\zeta$ at $r=0$ and the boundary condition at $r=R$ form a Sturm-Liouville problem. In such problems, the frequencies $\omega$ that solve the equation and satisfy the boundary conditions take discrete values. We place them in order as $\omega_0^{2} <\omega_1^{2}<\cdots<\omega_n^{2}$. For each eigenfrequency there is a corresponding eigenfunction $\xi_{n}$, which at order $n$ will have $n$ nodes. It is computationally more convenient to convert the second-order differential equation into a system of two first-order equations, which can be easily done by setting \(\eta=P\zeta'\)
\begin{align}
  \eta' &= - \left( Q+\omega^{2}W \right)\zeta\nonumber\\
  \zeta' &= \frac{\eta}{P}.
  \label{zeta_prime}
\end{align}
To find the solution we use the so called shooting method \cite{Kokkotas2001} that involves the following steps: i) we integrate the above equations starting  from the center of the star and stop at the surface for a trial value of $\omega^{2}$ satisfying the initial condition at $r=0$, ii) if at the surface the boundary condition $\eta(R)=0$ is satisfied, we have found one of the $\omega$, iii) if the boundary condition is not satisfied, we change the $\omega$ accordingly and we repeat the process until the boundary condition is satisfied. Once we have found an $\omega$ that satisfies the boundary condition, its order is determined by examining the number of nodes the corresponding eigenfunction $\xi$ possesses. Note that our equations and boundary conditions do not fix the overall amplitude of $\xi$, i.e., if $\xi_{n}$ is a solution, then $A\xi_{n}$ is also a solution. The overall amplitude is
fixed by the mechanism that creates the oscillation.
This gives us the freedom to choose $\eta(0)=1$. To avoid the aforementioned numerical issues at the origin of the star, we
Taylor expand $\zeta$ around $r=0$ 
\begin{equation}
\begin{split}
  \zeta(r)&=\zeta_{0}r^{3}+O(r^{5})\\
  \eta(r)&=1+O(r^{2}),
  \end{split}
  \label{Taylor_zeta}
\end{equation}
and integrate out not from $r=0$ but from an $r=r_0$ very close by. From the definition of $P$ (Eq.~(\ref{P})) and Eq.~(\ref{zeta_prime}), 
$\zeta_{0}=\left (3 \Gamma_1 p_0 e^{\lambda_0+3\nu_0} \right)^{-1}$. Since $\xi=e^{\nu_0}r^{-2}\zeta$, Eq.~(\ref{Taylor_zeta}) ensures the non-singular behavior of $\xi$ close to $r=0$. The same holds for $\delta p$ and the other thermodynamic perturbations.

Starting with a guess for $\omega$ which is close to the actual value reduces the number of iterations needed and speeds up the shooting process. To a good approximation the fundamental frequency should be of the order of the inverse of the free fall timescale
\[\omega \propto \sqrt{\frac{G M}{R^{3}}}.\]
After the first integration, we use the difference of $\Delta p(R)$ from $0$ to make a more educated guess for the new trial frequency. This is achieved by searching for the root of $\Delta p(R)$, using the secant method.
As a first check by choosing $z=107$ MeV and $m_{\chi}=1$ GeV, i.e., stars on the red line of the Fig.~\ref{fig:MR}, we verify a zero fundamental frequency for the star that corresponds to the maximum mass. This is anticipated since as we have already mentioned the solutions to the left of the maximum mass configuration are unstable, depicted by the fact that $\omega^{2}_{0}$ becomes negative, which subsequently means  an exponentially growing (and unstable) solution.

\begin{figure}[htbp!]
 \centering
 \includegraphics[scale=0.9]{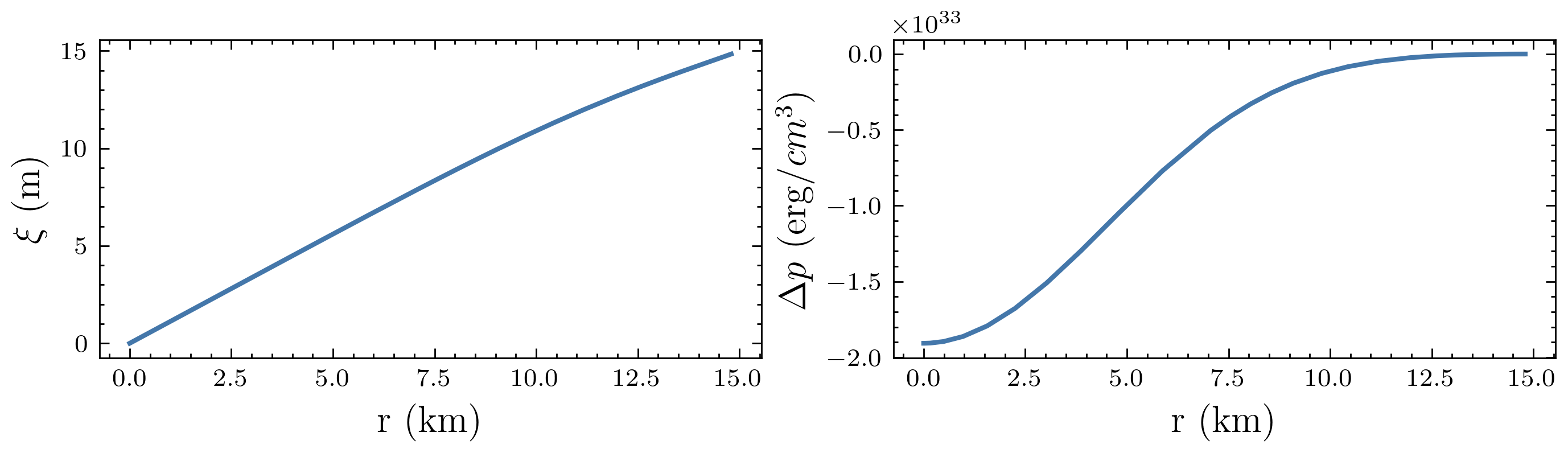}
 \caption{The radial displacement and the Lagrangian pressure variation \eqref{eq:lagpress} of the fundamental oscillation mode for the star with $z=107$ MeV and $\rho_{c}=1.5\times 10^{15} \text{gr}\;\text{cm}^{-3}$.\label{fig:O0} }
 \label{fig2}
\end{figure}

\begin{figure}[htbp!]
 \centering
 \includegraphics[scale=0.9]{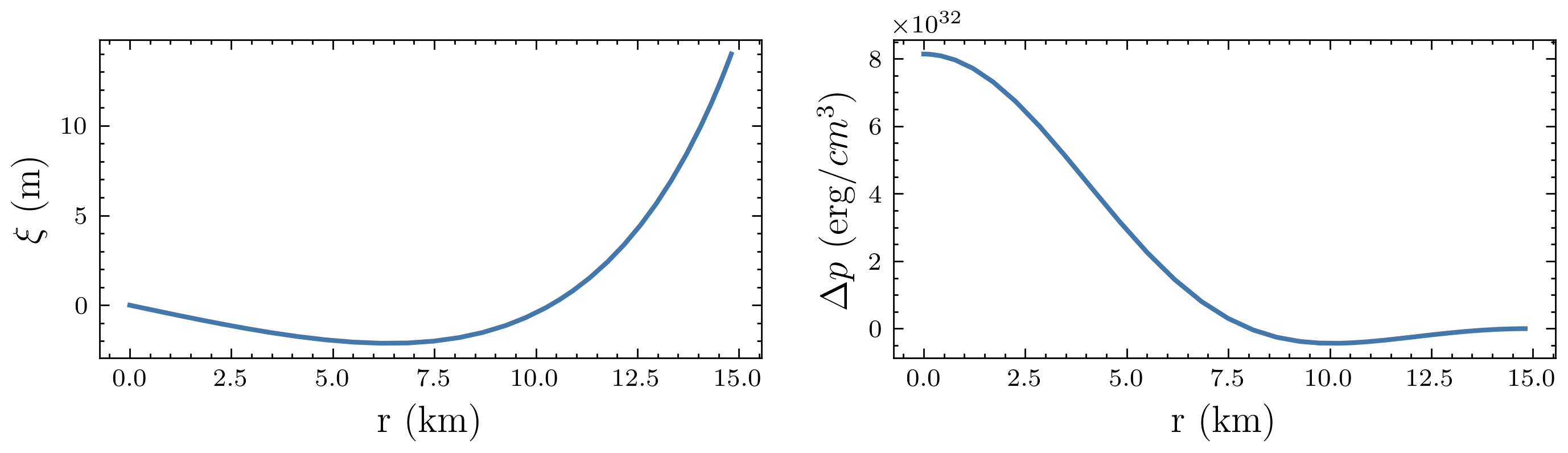}
 \caption{The radial displacement and the Lagrangian variation for the pressure of the first oscillation mode for the star with $z=107$ MeV and $\rho_{c}=1.5\times 10^{15} \text{gr}\;\text{cm}^{-3}$.\label{fig:O1}}
 \label{fig3}
\end{figure}

In principle the amplitude of the radial oscillation can be estimated once a known excitation mechanism is in place. However, this requires knowledge of all the dissipative mechanisms in place and an accurate estimate of the shear viscosity which goes beyond the scope of this paper. Therefore we  take the amplitude as an initial condition. In practice we make the reasonable assumption that the variation of the radius  is three orders of magnitude smaller than the radius itself, i.e.,  $\xi(R)/R=10^{-3}$.
In Figs. \ref{fig2} and \ref{fig3}, we present the variations corresponding to the fundamental and first oscillation modes of the star with central energy density $\rho_c=1.5 \times 10^{15} \text{gr}/\text{cm}^{3}$ and $z=107$ MeV. By inspection of Eqs. (\ref{eq:lagden}), (\ref{eq:lagpress}), and (\ref{eq:lagen}) it is evident that the three corresponding Lagrangian perturbations exhibit the same behavior as $\Delta p$. On the other hand, the Eulerian variations will be nonzero at the surface due to the term $- p'_0 \xi$, which becomes dominant at large $r$. Fig.~\ref{fig2} and \ref{fig3} depict the radial displacement and variation in the pressure as a function of $r$ for the fundamental and first oscillation mode respectively. Note the existence of a node in the first mode reflected in the fact that $\xi$ becomes zero for a radius different than $r=0$. Note also that the overall variation in pressure is significantly larger in the fundamental mode with respect to the first mode. This is a generic feature that as we will demonstrate  leads to distinct variations in the total number of photons received; higher order modes produce even smaller signals.

\section{Dark Matter Annihilation}\label{4}
Although we implement a model of asymmetric DM where there is a baryon-like conserved number for the DM particles, we assume that processes that slightly violate this number are in place. In particular we will assume that DM annihilation to two photons takes place. The annihilation cross sections we are interested in are tiny, i.e. suppressed by the Planck scale with typical values of the order of $\langle\sigma_{\rm ann } v \rangle =10^{-68}~\text{cm}^2$. Such cross sections  are much smaller than the DM freeze-out one by many orders of magnitude. With such a tiny cross section it is impossible to detect the  produced photons from the annihilation of free DM particles populating  the galaxy. However, since  the annihilation rate $\Gamma \sim n^2 \langle \sigma_{\rm ann} v\rangle$ depends crucially on the square of the DM density, even small annihilation cross sections can lead to large annihilation rates if DM particles populate a high density environment such as that of a compact dark star. Even tiny annihilation cross sections of the order of  $\sim 10^{-68}~\text{cm}^2$, can have  observable effects. For example let us compare the relative difference between the annihilation rate per volume by free DM particles of GeV mass in the galactic environment of a typical  galactic DM density of $\sim 1$ GeV$/\text{cm}^3$ and that of a compact dark star with a central density of $10^{14}~\text{gram}/\text{cm}^3$. Due to the $n^2$ dependence, there is a 75 orders of magnitude enhancement in the emitted rate for the latter case. 
Note that with such small annihilation cross section assumed here, one can easily prove that the overall reduction in the mass of the dark star takes place in timescales of billions of years. 
Dark matter particles with mass at $1\;\text{GeV}$ and a Yukawa interaction characterized by $z=107~\text{MeV}$ result in dark stars that exhibit similar parameters to those of typical neutron stars, as shown in Fig.~\ref{fig:MR}. Assuming a central energy density $\rho_c=1.5\times10^{15} ~\text{gr}/\text{cm}^3$, we get a star with $M=2 M_{\odot}$ and $R=15\text{km}$. A ballpark estimate of the star's luminosity can be obtained by assuming a uniform number density throughout the star given by $n=3M/(4 \pi R^{3}m_{\chi})$. The photon production rate is thus constant and can be trivially integrated over the star's volume yielding
\begin{equation}\label{eq:ballpark}
\frac{dN}{dt}=\langle\sigma_{\rm ann } v \rangle n^2 \frac{4}{3} \pi R^{3}=1.04\times 10^{38} s^{-1}. 
\end{equation}
A more accurate value of the rate can be obtained using the collision operator for $2\to2$ scattering.  In addition to providing a precise calculation, this approach allows us to obtain the energy spectrum of the emitted photons. The collision operator describing the number of photons produced in a volume $dV$ over a time period $dt$, is defined as
\begin{align}\label{eq:collisionoperator}
    \frac{C_{12\to 34}}{k_{12\to 34}}=\int &\frac{d^3p_1}{(2\pi)^32E_1}\int \frac{d^3p_2}{(2\pi)^32E_2}\int \frac{d^3p_3}{(2\pi)^32E_3}\int \frac{d^3p_4}{(2\pi)^32E_4}\times\nonumber\\
    &\times f_1 f_2 (1\pm f_3)(1\pm f_4) (2\pi)^4 \delta^4(p_1+p_2-p_3-p_4)|\mathcal{M}_{12\to 34}|^2,
\end{align}
where, $k$ is the symmetry factor, $f_i$ are the distribution functions describing each particle population and $\mathcal{M}_{12\to 34}$ denotes the matrix element of the interaction. Indices 1 and 2 correspond to DM particles and 3 and 4 to photons. Utilizing properties of the Dirac distribution we simplify the phase-space integrals. For convenience, we express the three-momenta in spherical coordinates and choose a reference frame in which one of the outgoing photons is aligned with the $z$-axis. Following \citep{Tuominen2022,Bringmann2023b} we parametrize the momenta as 
\begin{align}
\vec{p}_3 &= p_3(0,0,1),\nonumber\\
\vec{p}_4 &= p_4(\sin \beta,0,\cos \beta),\nonumber\\
\vec{p}_1 &= p_1(\sin \theta\cos \phi,\sin\theta \sin \phi ,\cos\theta).
\end{align}
We use the 3-dimensional part of the Dirac distribution to integrate over $\vec{p}_{2}$, imposing the constraint
\begin{equation*}
     {|{\vec{p}}_2|}^2=(\vec{p}_3+\vec{p}_4-\vec{p}_1)^2.
\end{equation*}
Using of the remaining $\delta$-distribution we fully integrate over the $\phi$ angle. Combing these steps leads to
\begin{align}
 \frac{C_{12\to 34}}{k_{12\to 34}}  = 
  \int& \frac{d^3p_3}{(2\pi)^32E_3}\int \frac{d^3p_4}{(2\pi)^32E_4}  \int dp_1 \frac{p^2_1}{(2\pi)^2E_1}\times\nonumber\\
  &\times f_1 f_2 (1\pm f_3)(1\pm f_4)\Theta(b^2-4ac)\int_{R_\theta} \frac{d\cos\theta}{\sqrt{a\cos^2\theta+b\cos\theta+c}}|\mathcal{M}_{12\to 34}|^2.
\end{align}
The parameters introduced are defined as
\begin{align}
&a=-4 p_{1}^{2} [(E_{3}+E_{4})^{2}-s],\nonumber\\
&b=\frac{2p_{1}}{p_{3}}[s-2E_{3}(E_{3}+E_{4})][s-2E_{1}(E_{3}+E_{4})],\nonumber\\
&c=-{[2E_1(E_3+E_4)-m_{1}^{2}+m_{2}^{2}-s]}^{2}- \frac{p_1^2}{p_3^2}s(s-s_{34,+}),
\end{align}
where $s$ is the Mandelstam variable and
\begin{align}
s_{12,+}&=2 m^2_x+2(E_1E_2+p_1p_2),\nonumber\\
s_{12,-}&=2 m^2_x+2(E_1E_2-p_1p_2),\nonumber\\
s_{34,+}&=4E_3E_4.
\end{align}
Having integrated out the Dirac distribution, the energy $E_2$ is given by energy conservation $E_2=E_3+E_4-E_1$ and thus $p_2=\sqrt{E^2_2-m^2_2}$.
The integration over $\cos\theta$ must be restricted to the set $R_\theta=\{-1 \le \cos\theta\le 1 |c_{\theta,+}\le \cos\theta\le c_{\theta,-}\}$ where
\[c_{\theta,\pm}=\frac{-b \pm \sqrt{b^2-4a c}}{2a}\] are the roots of the polynomial. A convenient change of variables is performed, replacing the remaining $\beta$ angle to $s$ using \[s=(p_3+p_4)^2=2\;E_3E_4-2\;\vec{p}_3 \cdot \vec{p}_4=2E_3E_4-2p_3p_4\cos\beta.\] The range of integration over $s$ becomes $R_s=\{s \in \mathbb{R}|\max(s_{12,-},s_{34,-}) \le s \le \min(s_{12,+},s_{34,+})\}$. With the above simplifications we express the collision operator regarding the number density of DM as
\begin{align}\label{eq:colop}
    C_{12\to 34}=&\frac{k_{12\to 34}\;g^2_s}{4(2\pi)^6}\int\limits_0^\infty dE_3\int\limits_{\max(0,2m_\chi-E_3)}^\infty \!\! dE_4 \!\!\int\limits_{m_\chi}^{E_3+E_4-m_\chi}\!\! dE_1\,p_1 \Theta(E_f-E_1)\Theta(E_f-E_2)\times\nonumber\\
  &\times\int_{R_{s}}ds\,\Theta(b^2-4a c)\int_{R_\theta}\frac{d\cos\theta}{\sqrt{a\cos^2\theta+b\cos\theta+c}}|\mathcal{M}_{12\to 34}|^2.
\end{align}
Here, the degeneracy factor $g_s=2s+1=2$ and the energy step functions come from the Fermi-Dirac distribution at $T \to 0$. The outgoing photons are not captured by the star and therefore do not have distribution functions.
To determine the energy spectrum of the emitted photons, we differentiate the collision operator with respect to $E_3$. We define this rate as $ \frac{dN}{dVdtdE_{3}}$ which simply gives the number of emitted photons with energy between $E_3$ and $E_3+dE_3$ per volume per time.
This photon production rate depends on $r$. This is because the DM number density is a function of $r$ and subsequently the Fermi momentum $p_F$ at a given point is also a function of r. Therefore $r$ not only affects the overall photon production but also the spectrum of the emitted energy because different layers have access to different maximum energy $\sqrt{p_F^2+m_{\chi}^2}$. 
Consequently, annihilation rates must be computed at different layers within the star, followed by integration over the star's volume to obtain the total contribution. $r$ affects the spectrum  in yet another nontrivial way.
Since we are interested in potential photon spectrum received on Earth based detectors, we must account for the redshift of the photon energies due to the gravitational potential of the star. Photons produced at different layers experience different redshifts. The relation between the emitted and the received energies is given by
\begin{equation}\label{eq:enredshift}
    E_{\text{obs}}=\sqrt{-g_{tt}}E_{\text{em}}=e^{\nu(r)}E_{\text{em}}.
\end{equation}
Therefore, we must account for the redshift while computing $ \frac{dN}{dVdtdE_{3}}(E_3,r)$. The calculation of the collision operator assumes a local rest frame in which $E_3$ is the energy of the emitted photon. The redshift affects the observed spectrum in  two ways: firstly the energy argument must be shifted from $E_{3}$ to $E_{\text{obs}}/\sqrt{-g_{tt}}$ and secondly, the differential energy element should transform as $dE_{\text{obs}}=\sqrt{-g_{tt}}dE_{3}$.
Due to the radial dependence of metric function $g_{tt}=-e^{2\nu}$, different layers of the star produce photons that experience varying degrees of redshift.
To obtain the observed spectrum we divide the star in a large number of shells with radius between $r$ and $r+dr$ and we compute
\begin{equation}\label{eq:ann_vol_en}
  \frac{dN}{dtdVdE_{\text{obs}}}\left(E_{\text{obs}},r\right)=\frac{1}{\sqrt{-g_{tt}}}\frac{dN}{dtdVdE_{3}}\left(\frac{E_{\text{obs}}}{\sqrt{-g_{tt}}},r\right),
\end{equation}
for a given shell and a bin of energy between $E_{\text{obs}}$ and $E_{\text{obs}}+dE_{\text{obs}}$, where  
\begin{equation}
    \frac{dN}{dtdVdE_{3}}=\frac{dC_{12\to34}}{dE_{3}} 
    \label{dNdVdt}
    \end{equation} 
    requires the evaluation of four integrals, as seen in (\ref{eq:colop}). Without loss of generality, we  assume  a constant matrix element for the interaction, something definitely valid in the case where the annihilation is mediated by a heavy e.g. boson. In that case the integral over $\cos\theta$ is evaluated analytically by changing variables from $\theta$ to f with \[f=-2 \arcsin{\left(\sqrt{(c_{\theta,-} - \cos\theta)/(c_{\theta,-} - c_{\theta,+})}\right)},\] which leads to the relation $\frac{df}{\sqrt{-a}}=\frac{d\cos\theta}{\sqrt{a\cos^{2}\theta+b\cos\theta+c}}$. The remaining integrations are performed numerically using an adaptive integration method.
This procedure allows us to evaluate \eqref{eq:ann_vol_en} at different radial shells within the star, yielding the number of annihilations per unit time and unit volume that produce photons with observed energies in the range \( [E_{\text{obs}},E_{\text{obs}}+dE_{\text{obs}}].\)\\
To obtain the observed photon spectrum, we perform a fifth integration over the star's proper volume
\begin{equation}\label{eq:vol}
  dV=\sqrt{g_{rr}}r^{2}\sin \theta d\theta d\phi dr=4\pi e^{\lambda(r)}r^{2}dr,
\end{equation}
where spherical symmetry has been used to integrate over the solid angle.  Integrating Eq.~(\ref{eq:ann_vol_en}) over $E_{\text{obs}}$ provides the number of annihilations from each star's shell. This will be used to compare the different approximations.
 The collision operator~\eqref{eq:colop} by definition measures the rate of change of the DM density. For every 2 DM that annihilate, 2 photons are produced, so the operator also measures the photon production rate. We calibrate the annihilation matrix element by setting $ C_{12\to 34} \equiv \langle \sigma v \rangle n^2$ and taking the zero velocity limit. Specifically in this limit by using the $\delta$ function and setting the energies equal to $m_\chi$ (ignoring the angle dependence of $p_3$) we can integrate out the $p_3,~p_4$ in ~\eqref{eq:collisionoperator}. Integrating over $p_1,~p_2$ with the Fermi-Dirac distribution \[n=\int \frac{d^3 p_i}{(2\pi)^3}f_i,\] gives the $n^2$ factor. Equating the result with $n^2 \langle \sigma v \rangle$ we get the matrix element 
\begin{equation}\label{eq:matrixelement}
  {|\mathcal{M}_{12\to34}|}^2 = \frac{16\pi}{k_{12\to 34}}  \; m^{2}_{\chi} \langle\sigma v \rangle .
\end{equation}
In our  calculation of the successive four numerical integrations required, we have set  the relative error tolerance at $10^{-7}$.
This requirement ensures that numerical errors in the innermost integrations do not accumulate significantly as subsequent integrations are performed. For the grid points $(E_{obs},r)$ we choose $E_{\text{obs,points}}=400$ points for $E_{obs}$ and divide the star in $r_{\text{points}}=160$ shells. This partitioning is done by taking into account the computational time and the accuracy benefits of a more dense grid.  Our results have remained robust within all the  runs with different number of grid points we tried. For example we observe no noticeable differences in the energy spectrum for $r_{\text{points}}=100$ and $200$, while $dN/dt$ varies only at the third decimal point.

The presented calculation gives accurately the photon spectrum and luminosity produced by the DM annihilation. However it is worth comparing the result to naive  expectations where all the kinematics are ignored. For example, one can compare the total number of photons per time produced by the annihilation regardless of energy. Within our frame, we can do that  by integrating Eq.~(\ref{eq:ann_vol_en}) over both the unit volume (\ref{eq:vol}) and the energy $E_{\text{obs}}$.
For one of the benchmark dark stars we studied i.e., $z=107 ~\MeV$, $\rho_c=1.5\times 10^{15}\; \text{gr}/\text{cm}^3$ and $\langle\sigma v\rangle=10^{-68}\text{cm}^{2}$, the total photon production rate is
\begin{equation}
\frac{dN}{dt}=0.9\times 10^{38} s^{-1}.
\label{dNdtr}
\end{equation}
This result is very close to the naive ballpark estimate of  Eq.~(\ref{eq:ballpark}), where all kinematics were ignored and the star was treated as an object with uniform density. Since the overall annihilation rate is $\sim n^2$, a constant density approximation underestimates the annihilation at the core and overestimates the annihilation in the outer layers. The dominant contribution to the annihilation rate comes from shells that are underestimated by the constant density approximation and therefore one would naively expect the rate of Eq.~(\ref{eq:ballpark}) to be smaller than that of Eq.~(\ref{dNdtr}). However, the kinematic restrictions which are taken into account in the latter make the rate to be slightly smaller than the former where no kinematic considerations have been included.

\begin{figure}[!t]
  \centering
  \includegraphics{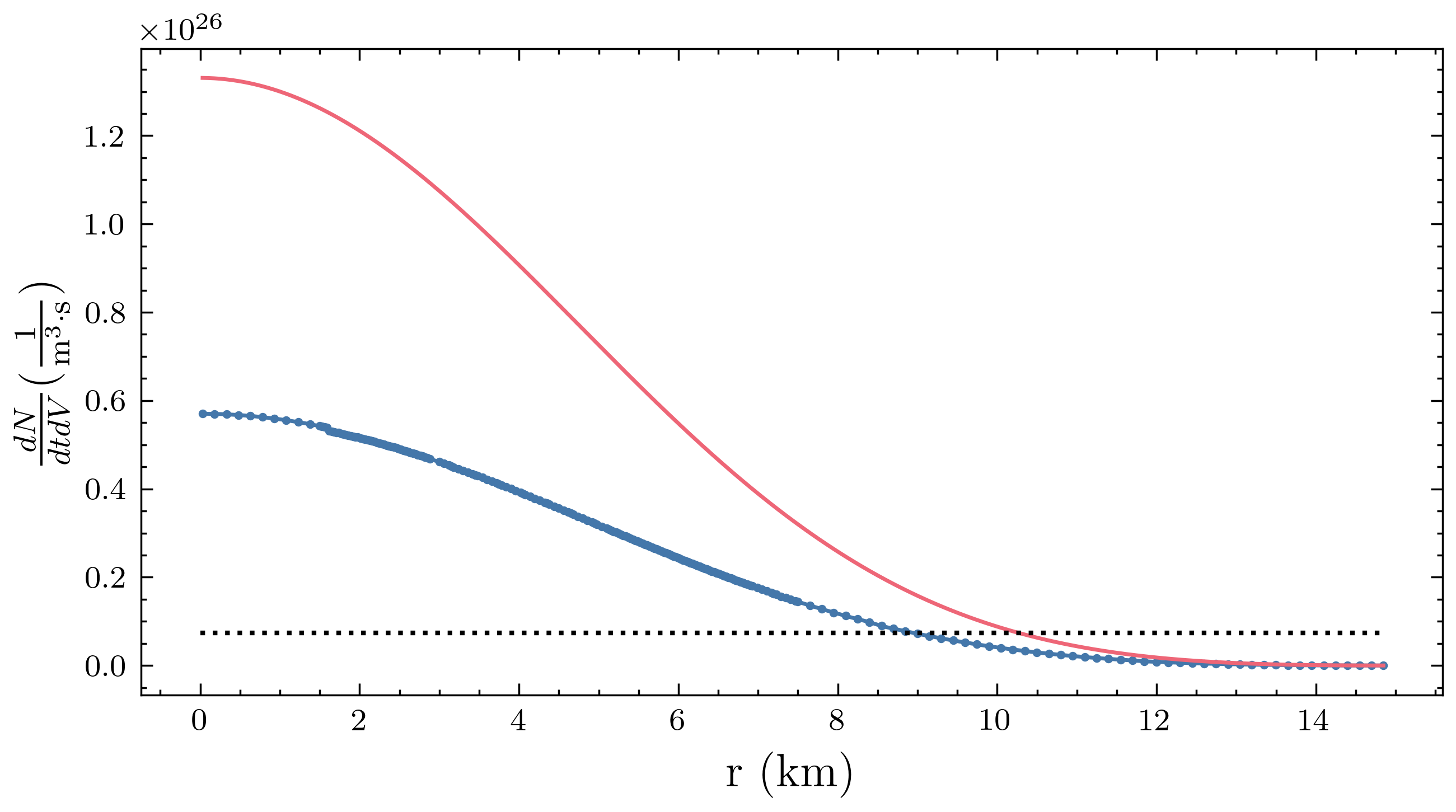}
  \caption{\label{fig:annpershell} Photon production per unit time and unit volume as a function of the layer's location of the star. The blue graph shows the full numerical computation presented, the red one is the approximation $n^2\langle \sigma v \rangle$ and the dotted line displays the same quantity under the $n=\text{const.}$ approximation.}
\end{figure}

Another approximation for the photon production rate can be obtained by taking into account the actual density profile of the star, i.e., computing
\begin{equation*}
\frac{dN}{dtdV}=n^2(r) \langle \sigma v \rangle.
\end{equation*}
The total photon production rate is subsequently given by
\begin{equation}\label{eq:approxann}
  \frac{dN}{dt}=\int\limits_{0}^{R} dr\;\; 4 \pi e^{\lambda(r)}r^{2}n^2(r) \langle \sigma v \rangle,
\end{equation}
which for the previous example yields
\[\frac{dN}{dt}= 2\times  10^{38}\;s^{-1}.\]
We should note that this approximation gives a photon production rate roughly twice that  of the full numerical computation of Eq.~(\ref{dNdtr}). This approximation overestimates the actual number of annihilations, a discrepancy that arises because the kinematical constraints are ignored under this approximation. This rate is also larger than the constant density approximation too for two reasons. Firstly as we have already mentioned, the constant density underestimates the overall photon production rate and secondly the relativistic correction of the volume integral of Eq.~(\ref{eq:approxann}) give effectively a larger volume from the naive nonrelativistic one. 
Fig.~\ref{fig:annpershell} depicts the photon production rate per volume of the three aforementioned estimates. These rates  differ more in the inner layers agreeing with the general expectations. Inner layers correspond to higher densities and therefore higher Fermi momenta. Satisfying energy and momentum conservation in these cases is more restricted than the regions with smaller momentum. The estimate where the density is considered constant throughout the star seems to be significantly smaller. Despite these differences, as we have pointed out above, the overall rates once one integrates over the whole star are not larger than $\sim 2$. This is because inner layers contribute parametrically with smaller volumes, while layers with low depth where the rates are not different contribute more due to larger volumes.

Using Eqs.~(\ref{eq:colop}), (\ref{eq:ann_vol_en}), (\ref{dNdVdt}) we can find the photon emission (including the redshift) for each layer of the star. By dividing the star in a large number of layers and adding up their contributions, we can find the total spectrum of emitted photons. We have chosen four benchmark sets of parameters which correspond to four different dark stars (with repulsive Yukawa interactions) by fixing $z$, and $\rho_c$. For all the cases we have taken $m_{\chi}=1$ GeV. Table~\ref{tab:gaussian_fit} depicts the parameters used. It is worth noting that the last two stars have the same $z$ value but they differ in the central density $\rho_c$. They belong to the same family of solutions shown (red curve) in Fig.~\ref{fig:MR} and represented by the two dots on it. Note also that the first three stars represented in the table have core densities slightly less than the maximum values presented in Fig.~\ref{fig:MR}. We should also stress here the following pattern: stronger repulsive Yukawa interactions correspond to smaller $z$ and lead to larger maximum masses with smaller core densities. This is understood by the fact that Yukawa interactions provide a significant fraction of the pressure for the system and therefore there is no need for large densities and thus large Fermi pressure to sustain the star. Since the annihilation rate depends strongly on the density, it is no surprise that dark stars with weaker Yukawa interactions produce larger luminosity. In fact we have found that the produced photon spectrum in all cases is well fitted 
by  Gaussian distributions. The parameters for the Gaussian fit of each presented dark star are reported in Table~\ref{tab:gaussian_fit}, along with the mean photon energy after taking into account the redshift. 
\begin{table}[hbpt!]
\centering
\resizebox{\textwidth}{!}{\begin{tabular}{|c|c|c|c|c|c|c| }
  \hline
\(z\) (MeV)&$\rho_c$ $(10^{14}\text{gr}/\text{cm}^3)$&$M\;(M_\odot$)&$R$ (km) & $dN/dt$ $(10^{37}/s)$  & Mean Energy (MeV) & \(\sigma\) (MeV) \\
\hline
71 &7.4 &2.84&21.1& 6.7 & 605& 87\\
89 &11.3&2.32&17.3 & 8.2 & 611& 94\\
107&15.5&1.98&14.9 & 9.1 & 620& 100\\
107&3.0 &1.43&21.5& 1.7 & 837& 74\\
\hline
\end{tabular}} 
\caption{Parameters for the four dark stars studied, including the mean energy and standard deviation of the Gaussian fit.}
\label{tab:gaussian_fit}
\end{table}

\begin{figure}[t]
    \centering
    \includegraphics[]{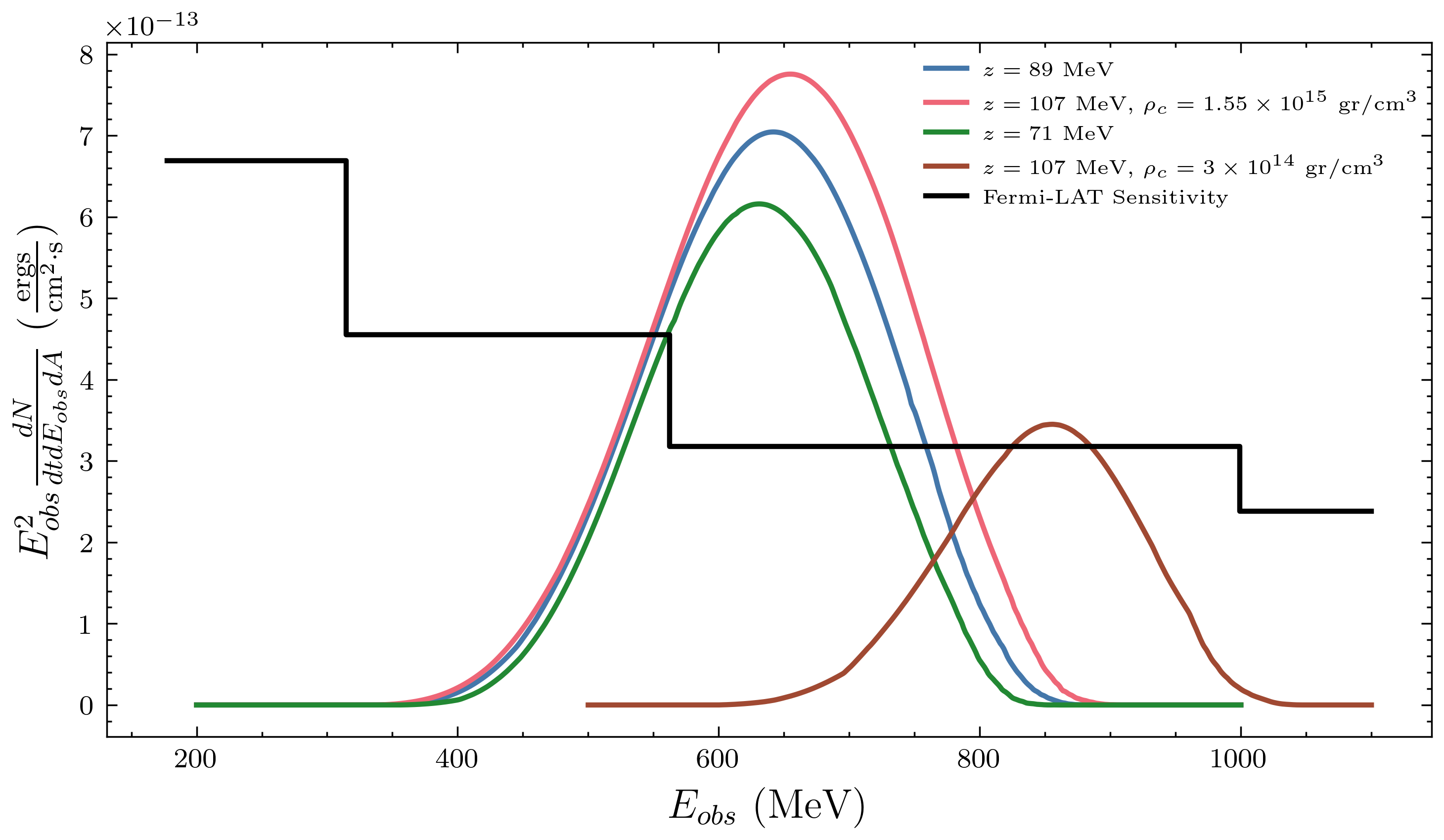}
    \caption{Differential sensitivity for the four benchmark stars with $\langle \sigma v \rangle =10^{-68} \text{cm}^{2}$ and $d=50$ kpc. Each colored line corresponds to a different star from Table~\ref{tab:gaussian_fit}. The black line is the Fermi-LAT sensitivity \citep{FermiLAT} in this energy range.}
    \label{fig:DifferentialSensitivity}
\end{figure}

Once the produced photon spectrum  has been derived, one can estimate the differential photon flux arriving on Earth based detectors. This is given by
\begin{equation*}
  E^{2}_{obs} \dfrac{dN}{dtdE_{obs}dA} = \frac{E^{2}_{obs}}{4\pi d^{2}} \int\limits_0^R dr 4\pi r^{2} e^{\lambda}\frac{dN}{dtdVdE_{obs}},
\end{equation*}
where $d$ is the distance to the dark star and $dA$ is the unit surface. Fig.~\ref{fig:DifferentialSensitivity} shows the differential photon flux as a function of the photon energy for the four reference stars of Table~\ref{tab:gaussian_fit}, located at a hypothetical distance of $d=50$ kpc with annihilation cross section $\langle \sigma v \rangle =10^{-68}~\text{cm}^2$ and contrasted to
 Fermi-LAT's current sensitivity~\cite{FermiLAT}. The differential sensitivity of Fermi-LAT at $E \approx 600\; \MeV$ is \( 3\times 10^{-13} {\text{ergs}}/{\text{cm}^{2}\text{s}}\). 
 For the studied stars of Table~\ref{tab:gaussian_fit} with the aforementioned distance of $d=50\; \text{kpc}$ from the Earth, we find the observed  flux to be  above the observational limit. 
  The flux  is directly proportional to the ratio $\langle \sigma v\rangle/d^{2}$. Smaller annihilation cross sections can still give the same flux if they are counterbalanced by a smaller distance. 

\section{Luminosity Modulation}\label{5}
In sec.~\ref{3} we studied the radial oscillations of dark stars provided an excitation mechanism is present. If the stars are prone to such oscillations, the received photon spectrum is expected to vary over time. 
Given the small amplitude of oscillations, we assume that the shape of the photon spectrum remains largely unchanged, with the oscillations primarily affecting the overall luminosity that will be modulated. We will comment later on the changes in the spectrum. We make this approximation because it becomes computationally very expensive to follow the procedure of the previous section for different  snapshots of the oscillating star.   
Instead it is much easier to estimate the overall variation in the total photon production due to the radial oscillation.  In particular, the oscillation modifies the number density as $n(r)\to n(r)+\delta n(r)$ while the volume element changes in two ways. Firstly, the perturbation of the metric function $\lambda \to \lambda +\delta \lambda$ varies the proper volume and secondly, the direct change in radius $R \to R + \xi(R)$ gives rise to a surface term. Thus, the variation of the photon production rate is
\begin{equation}\label{eq:oscillatingann}
\frac{dN}{dt}= 4\pi \langle \sigma v \rangle \int\limits_{0}^{R+\xi(R)}\! dr\;r^{2}\; e^{\lambda_{0}(r)+\delta\lambda(r)} {\left[  n_{0}(r)+\delta n(r) \right]}^{2}.
\end{equation}
Since these perturbations are small  we can Taylor expand the above integral. Keeping only up to first order terms we obtain
\begin{align}\label{eq:perturbedsignal}
 \frac{dN}{dt}&= 4\pi \langle \sigma v \rangle \int\limits_{0}^{R}\! dr\; r^{2}e^{\lambda_{0}(r)}\left(1+\delta\lambda(r) \right) n^{2}_{0} {\left( 1+2 \frac{\delta n(r)}{n_{0}} \right)}+ 4\pi \langle \sigma v \rangle \!\int\limits_{R}^{R+\xi(R)}\! dr\;r^{2}\; e^{\lambda_{0}(r)} { n_{0}(r)}^{2}\nonumber\\
  &\begin{aligned}
    =\left.\frac{dN}{dt}\right\rvert_{\text{un}}+4\pi \langle \sigma v \rangle \Bigg( \int\limits_{0}^{R}\! dr\; r^{2}e^{\lambda_{0}(r)}n^{2}_{0} \delta\lambda(r)+&\int\limits_{0}^{R}\! dr\; r^{2}e^{\lambda_{0}(r)}n_{0}\;\; 2 \delta n(r)\\
  &+e^{\lambda_{0}(R)}R^{2} n^{2}_{0}(R) \xi(R) \Bigg),
    \end{aligned}
\end{align}
where
\begin{equation}
\left.\frac{dN}{dt}\right\rvert_{\text{un}}=4\pi \langle \sigma v \rangle \int\limits_{0}^{R}\! dr\; r^{2}e^{\lambda_{0}(r)}n^{2}_{0}
\end{equation}
is the unperturbed photon production we  calculated in the previous section.
Each perturbation term carries a common time-dependent factor $e^{-i \omega t}$, where $\omega$ is the oscillation frequency. All the quantities necessary for the perturbation calculations are computed according to the radial oscillation section. 
Applying \eqref{eq:perturbedsignal} to the benchmark star with $z=107\; \text{MeV}$ and $\rho_c=1.5\times 10^{15}~ \text{gr}/\text{cm}^3$ and assuming an oscillation amplitude $\xi (R)/R\sim 10^{-3}$, we obtain a modulation of the photon production by roughly two orders of magnitude smaller than the unperturbed rate.
The frequency of the modulation is that of the fundamental mode $f_{0}=\omega_0/2\pi= 0.076~\text{kHz}$. It is useful to estimate the effects of oscillations when considering higher modes. Their contribution to the luminosity modulations becomes progressively smaller. In the previous mentioned benchmark star, the first excited mode has a frequency \(f_1= 3.5 \; \text{kHz}\) (which is much larger than the fundamental one)  and the corresponding variation to the photon production rate becomes roughly four orders of magnitude smaller than the unperturbed rate.
 This pattern continues for even higher-order modes, with their contributions decreasing further.

\begin{figure}[!tbp]
    \centering
    \includegraphics[]{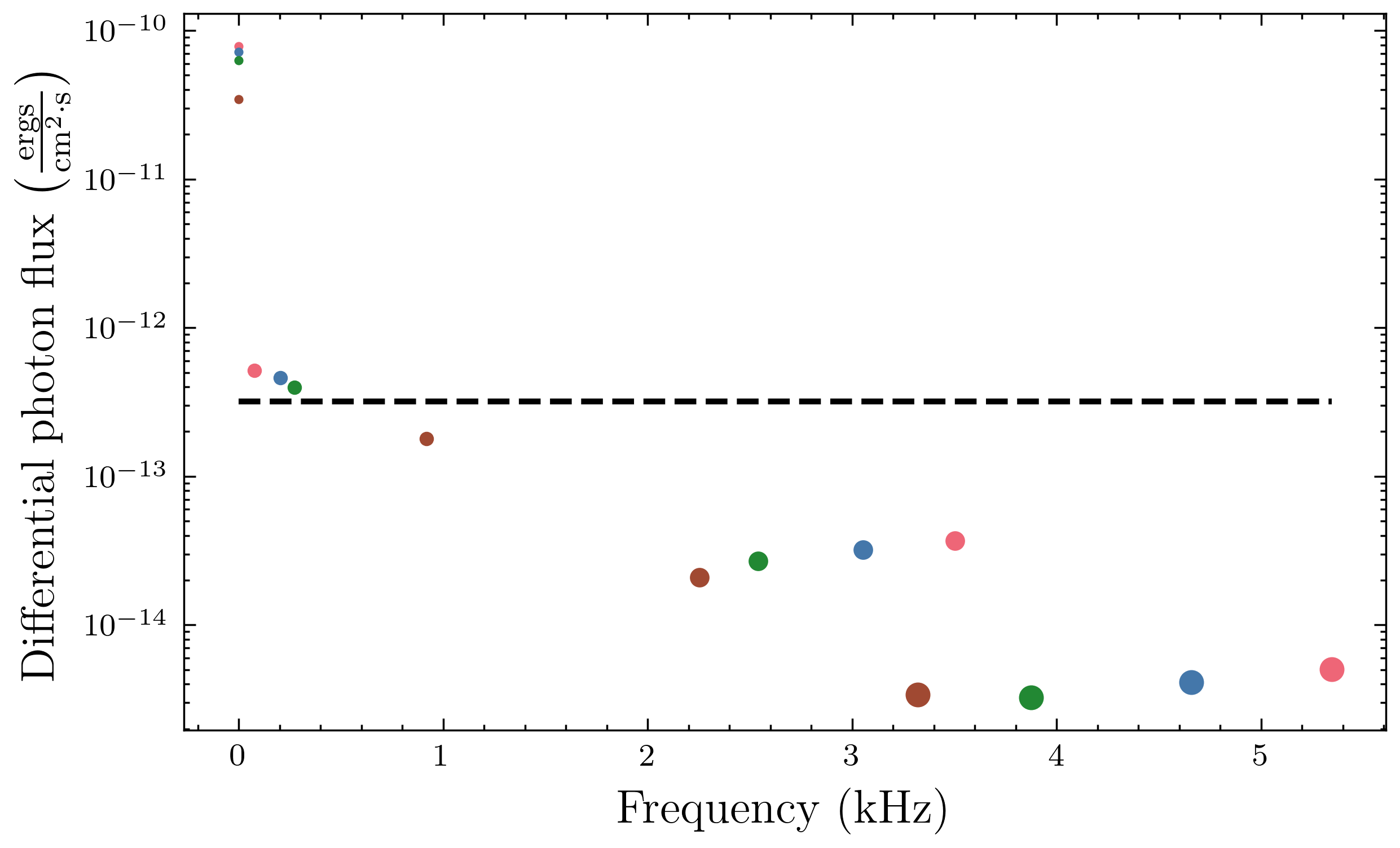}
    \caption{Maximum differential flux of the first three oscillation modes for the four benchmark stars as a function of the modulation frequency. The colored dots represent different stars, with the same colors as in previous plots. The dots with the smallest size correspond to the unperturbed flux. Different sizes of the dots correspond to different oscillation modes with a larger size depicting a higher mode.  
    The dashed line is the Fermi-LAT sensitivity in this energy range.}
    \label{fig:OscSens}
\end{figure}

By following the same procedure as with the unperturbed signal, we obtain the differential flux for the oscillation modes. We  obtain the peak differential flux of the perturbed spectrum by assuming that the shape of the spectrum does not change. 
In Fig.~\ref{fig:OscSens} we show the differential photon flux (contrasted against the Fermi-LAT sensitivity) for the four benchmark stars of 
Table~\ref{tab:gaussian_fit} as a function of the luminosity modulation frequency for $\langle\sigma v\rangle =10^{-66}\;\text{cm}^{2}$. Each dot represents a star in a particular mode. The four vertical dots on the upper left corner correspond to the constant unperturbed flux which trivially have zero frequency. The next tetrad of points  to the right and lower correspond to the fundamental mode for the same four benchmark stars. Moving to the right and lower we depict the next two oscillation modes with respect to their corresponding frequencies. As anticipated, each higher mode corresponds to a larger frequency and a smaller amplitude compared to the previous mode.
As it can be seen from the plot,
 the fundamental mode can have frequencies ranging from 10 Hz to 1 kHz. Generally, stars away from the maximum mass limit have higher fundamental frequencies. This can be verified by comparing the two stars (red and brown dots) with the same $z$ but different central density. The star with the smaller $\rho_c$ (brown dot) is further from the TOV limit and its fundamental frequency is higher, while its overall amplitude lower. The extreme case occurs exactly at the maximum mass, where the frequencies become imaginary, indicating the onset of the instability.

As we mentioned at the beginning of this section, we have assumed that although the radial oscillations alter the luminosity of the dark stars, the frequency spectrum remains unaffected in shape. We now demonstrate why this is the case. 
\begin{figure}[!t]
    \centering
    \includegraphics[]{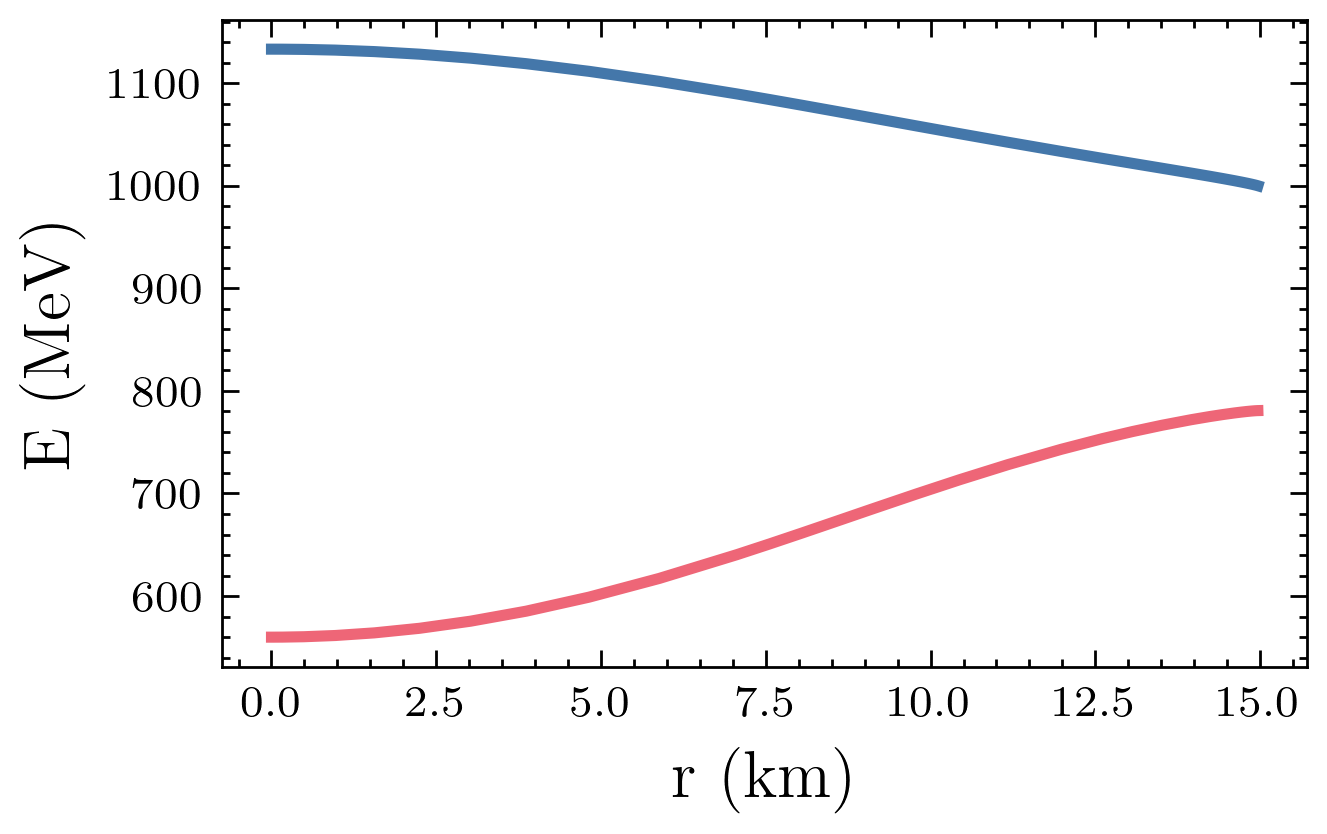}
    \caption{Fermi energy $E_{F}=\sqrt{m^2+p^2_{F}}$, for the star with $z=107\; \MeV,\; \rho_c =1.55\times10^{15} \text{gr}/\text{cm}^3$, as a function of $r$ (blue line). The red line is the red-shifted Fermi energy as given by Eq.~(\ref{eq:enredshift}).}
    \label{fig:redshift}
\end{figure}
 The perturbations of the number density affect the Fermi energies of the different layers, which should alter the standard deviation $\sigma$ of the spectrum in Table~\ref{tab:gaussian_fit}. We can approximate this effect by analyzing the two ends of the spectrum. The lower-energy photons originate from the center of the star, while higher-energy ones come from the surface. This occurs due to the redshift being higher at the center, than  the surface. As shown in Fig.~\ref{fig:redshift}, the higher redshift at the center is more than enough to counterbalance the larger Fermi energy at the center. 
 The Fermi energy at each layer dictates the maximum energy for the annihilation and thus the maximum energy of the photons. From Eq.~(\ref{numberd}) we see that
\begin{equation}
        x =\frac{p_{F}}{m_{\chi}}= \frac{(3\pi^2)^{(1/3)}}{m_{\chi}}n^{1/3},
\end{equation}
leading to 
\begin{equation}
    \frac{\delta x}{x}= \frac{1}{3} \frac{\delta n}{n} \sim 10^{-3}.
\end{equation}
We can relate this to the perturbation in the Fermi energy
\begin{equation}
    E_{obs}=\sqrt{-g_{tt}} \left(m_{\chi}\sqrt{1+x^2}\right)=e^{\nu} \left(m_{\chi}\sqrt{1+x^2}\right) \Rightarrow 
\frac{\delta E_{obs}}{E_{obs}}=\delta\nu+\left(\frac{x^2}{1+x^2}\right)\frac{\delta n}{3\;n}.
\end{equation}
The factor in the parenthesis ranges from 0.2 at the center to zero at the surface, while $\delta n/n \sim 10^{-3} $. The variation of the metric function, $\delta \nu$, is also of order $10^{-3}$ at the center and zero at the surface. Interestingly, when $\delta n<0$, we have $\delta \nu > 0 $ suggesting that the two effects tend to balance each other out.   
Consequently, the maximum received energy does not change to first approximation while the minimum energy experiences a relative change of the order of $10^{-4}$. Therefore, we can safely conclude that the perturbed signal follows a Gaussian with the same mean energy and standard deviation, as the background unperturbed signal. The integral of the spectrum is set equal to the $dN/dt$ perturbation of each mode, calculated from (\ref{eq:perturbedsignal}).

\section{{Dark Star Pulsation: \Large $\epsilon$}-Mechanism}\label{6}
In this section we present the  conditions under which the dark stars can become prone to sustainable radial oscillations.
Firstly, encounters of the dark star with other massive objects such as black holes and/or neutron stars can excite non-radial modes that through nonlinear effects can excite also a radial component. In principle,
shear viscosity, as well as thermal conductivity can damp an oscillation once such a mode is excited. How fast the attenuation of the oscillation occurs depends on the strength of the dissipative forces. However as it is well known from ordinary variable stars like e.g. the Cepheid variables, there can be mechanisms in place that feed the oscillation with energy, prohibiting the oscillation from attenuating with time. Examples of this kind in ordinary stars are the $\kappa$- and $\epsilon$-mechanism. In the $\kappa$-mechanism the opacity of the outer layers increases with increasing temperature. This is in contrast to the typical situation where opacity follows Kramers' law where $\kappa \sim T^{-5/2}$. As it was pointed out firstly by Eddington, the whole process works as a valve. As the star contracts, temperature rises which subsequently increases also the opacity, thus impeding heat flow. This obstructed heat flow builds up until eventually pushes the fluid shell further than the equilibrium position. At this new position temperature and density drop causing also the opacity to drop. The heat evades easier and the shell loses support and starts contracting again. 

In the case of $\epsilon$-mechanism a similar behavior occurs via the energy production from fusion per time per mass $\epsilon$. This is a quantity that depends on the nature of fusion nuclear reactions where within a range of parameters $\epsilon \sim T^a$ where $a$ is a positive power. As the star contracts, the temperature and density rise, increasing $\epsilon$ and subsequently the luminosity of the layer. The increased radiation pressure pushes the layer outward. As the star expands, the temperature and density drop and therefore so do $\epsilon$ and luminosity. As the radiation pressure also drops the layer loses support and start contracting again. For typical stars the $\epsilon$-mechanism is not effective. This is because it is only in the inner core of the star where fusion and energy production takes place. Dissipation of heat in the outer layers where there is no energy production can damp the oscillation. In such a case there is no sustainable oscillation through that mechanism. However, in supermassive stars where radiation pressure plays an important role, the $\epsilon$-mechanism can be in place~\cite{Ledoux1941}.

During an oscillation, the different layers of the star contract and expand, returning to the same position after the completion of one period. If the net work $W$ done by all shells on their surroundings is positive, the oscillation mode will be sustained. If $W<0$, the oscillation is doomed to die out~\cite{Cox74}. When there is a lag in the pressure variation with respect to the density one, $W>0$ while in the case of a lead of the former compared to the latter $W<0$. 

In our case an $\epsilon$-mechanism can feed the cycle of the pulsation with $W>0$ with the difference  to the case of ordinary stars being that the production of energy is not coming from fusion but from DM annihilation. In the case we have considered i.e., slight DM annihilation, the energy production as we have already mentioned is $dE/(dVdt)=(\rho^2/m_{\chi})\langle \sigma_{\rm ann} v\rangle$ which can be rewritten as $\epsilon=(\rho/m_{\chi})\langle \sigma_{\rm ann} v\rangle$.
Despite the fact that the DM is degenerate, there is nevertheless a nonzero temperature which can affect the system in two different ways. Firstly, there can be a small but nonzero contribution to pressure from radiation. In our model we have assumed DM annihilation to SM particles but one can envision also a partial annihilation to light particles of the dark sector. The second way that the temperature affects the system is due to the fact that non-interacting Fermi gas has a temperature dependent component which for $T<<p_F$ is $p=p_0+\pi^2 \rho^2 T^2/(15 m_{\chi}^2 p_0)$ where $p_0$ is the Fermi pressure at $T=0$. The instability to pulsation is quite similar to ordinary supermassive stars where the $\epsilon$-mechanism could be in play. As the star contracts, the density and temperature increase leading to a higher DM annihilation rate which increase in turn the pressure and drives the expansion of the star's layers. At maximum expansion, the lower density and temperature reduce subsequently the pressure and the layers lose support and contract. This is quite similar in act as a thermodynamic Carnot engine that produces positive work in each cycle.
Therefore this physical mechanism drives the pulsation as long as the net work $W$ done by all the layers is positive or as we have discussed, it suffices to show that there is a lag in pressure variation with respect to the density one. To demonstrate this, we linearize the first law of thermodynamics which takes the form~\cite{Cox74}
\begin{equation}
    \frac{\partial}{\partial t}\left (\frac{\Delta p}{p} \right )=\Gamma_1\frac{\partial}{\partial t}\left (\frac{\Delta n}{n} \right )+\frac{n (\Gamma_3-1)}{p}\Delta \left (\epsilon -\frac{\partial L}{\partial m} \right ),    
    \label{1stLT}
    \end{equation}
where $\Gamma_3-1 \equiv \left (\frac{\partial \ln T}{\partial \ln n} \right )_s$ ($s$ stands for adiabatic process). $\Delta$ denotes variation from the equilibrium value and  $L$ is the luminosity as a function of the included mass $m$. Note that this luminosity does not include the photon emission since photons once produced evade from the star, but rather only the component related to emission of particles of the dark sector. The same holds for $\epsilon$, i.e., 
\begin{equation}
\epsilon=f(\rho/m_{\chi})\langle \sigma_{\rm ann} v\rangle,
\label{epsilon_DP}
\end{equation} 
$f$ being the fraction of the annihilation that goes to the dark sector. A comment is in order here. Ignoring the second term in the right hand side of Eq.~(\ref{1stLT}) corresponds to an adiabatic process. The second term is the heat input into the system that violates adiabaticity. In our study of the radial oscillations we assumed the process to be quasi-adiabatic. Therefore it is important to check in fact if our approximation is valid, i.e., if the second term of Eq.~(\ref{1stLT}) is much smaller than the first. Practically speaking this is equivalent to demonstrating that the Kelvin timescale $t_K=GM^2/(RL)>>t_{ff}$. Given the fact that we know the energy production rate of DM annihilations, we have checked that for our cases of interest,  the Kelvin timescale is many orders of magnitude larger than  the free fall one and therefore the process is indeed quasi-adiabatic.

At the point of maximum compression $\partial \Delta n/\partial t =0$ in Eq.~(\ref{1stLT}), one can realize that as long as $\Delta (\epsilon -\partial L/\partial m)>0$, then the pressure variation lags relatively to the density variation and therefore we have positive work done by a star's layer. If the net result of all layers is the same, it means that the dark star is unstable to pulsations. From Eq.~(\ref{epsilon_DP}) and for an s-wave annihilation (i.e., $\langle\sigma_{\rm ann} v\rangle$ is constant) we see that $\epsilon \sim \rho$ and therefore $\Delta \epsilon/\epsilon =\Delta \rho /\rho$. If for a moment we ignore $\partial L/\partial m$, it is clear why there is an $\epsilon$-mechanism in play here. At maximum contraction according to Eq.~(\ref{1stLT}) $\partial \Delta p/\partial t\sim \Delta \rho>0$ and therefore the pressure is still increasing and $W>0$. 

Although an $\epsilon$-mechanism can be in place here, one has to take into account the leakage of energy via transport which tends to undo the effect of the $\epsilon$ term, i.e., we must consider the effect of $\partial L/\partial m$. To this end, the net work done by all the layers of the star over a cycle of the oscillation of period $T$ is
\begin{equation}
W=\int_0^M dm\oint p dV=\int_0^M dm\int_0^T p\frac{\partial}{\partial t} \left (\frac{1}{n}\right )dt.
\end{equation}
The sign of $W$ determines whether an instability for oscillation will be set off or not. Given that at the end of a cycle the internal energy of the system returns to its initial value, the work done will be equal to the net heat absorbed by all the layers. Therefore the above equation can be recast as
\begin{equation}
W=\int_0^M dm\int_0^T (\Gamma_3 -1)\frac{\Delta n}{n} \Delta \left (\epsilon -\frac{\partial L}{\partial m}\right ) dt.
\label{workdone}
\end{equation}
The variation of energy produced by annihilation stuck inside the star is in general
\begin{equation}
\frac{\Delta \epsilon}{\epsilon} = \frac{\Delta n}{n}\left[\left(1+\frac{P}{\rho}\right)\epsilon_\rho +\epsilon_T (\Gamma_3-1)\right],
\label{deltaE}
\end{equation}
where $\epsilon_\rho \equiv d\ln\epsilon/ d\ln\rho|_T$, $\epsilon_T \equiv d\ln\epsilon/ d\ln T|_\rho$ and $\epsilon$ is in the equilibrium configuration. From the  discussion above is evident that for an s-wave annihilation $\epsilon_\rho=1$ and $\epsilon_T=0$.  In the case of p-wave and d-wave annihilation  $\langle \sigma_{\rm ann}v \rangle$ scales as $v^2$ and $v^4$ respectively. Since we assume that DM particles form a degenerate fluid with a temperature smaller than the chemical potential, $v$ can be effectively replaced by the Fermi velocity $v_F \sim p_F\sim \rho^{1/3}$. Therefore for p-wave and d-wave annihilation $\epsilon_\rho =1+(2/3)=5/3$ and $\epsilon_\rho =1+(4/3)=7/3$ respectively with $\epsilon_T=0$ for both cases.

For the luminosity term of Eq.~(\ref{workdone}), we have
\begin{equation}
    \Delta\left (\frac{\partial L}{\partial m}\right )=
\frac{\partial}{\partial m}\left (L\frac{\Delta L}{L}\right )=\epsilon \frac{\Delta L}{L}+L \frac{\partial}{\partial m}\left (\frac{\Delta L}{L}\right ).
\label{DeltaLm}
\end{equation}
Recall that for each layer of the star we have $\partial L/\partial m=\epsilon$. The transport  of energy inside the star can take place in general via three different ways: radiation, convection and conduction. We will ignore convection here and consider primarily radiation. As aforementioned, the annihilation of DM can have a branch to photons as we have studied in the previous section and another branch to e.g. some light scalars of the dark sector that provide radiation (and in particular radiation pressure). At high density conduction can change the form of opacity and in general one has to take $\kappa^{-1}=\kappa_{r}^{-1} + \kappa_{\rm cd}^{-1}$ where the three $\kappa$ are the total opacity, radiation opacity and conductive opacity respectively. Recall that temperature and luminosity are related via
\begin{equation}\label{eq:tempm}
    \frac{\partial T}{\partial m}=-\frac{3}{64 \pi^2 a}\frac{\kappa L}{r^4 T^3},
\end{equation}
where $a$ is the radiation density constant. Using this equation we get
\begin{equation}
\frac{\Delta L}{L}=4\frac{\delta r}{r}+\frac{\Delta n}{n}[(4-\kappa_T)(\Gamma_3-1)-\left(1+\frac{P}{\rho}\right)\kappa_{\rho}]+\frac{1}{d\ln T/dm}\frac{\partial}{\partial m}\left [\frac{\Delta n}{n}(\Gamma_3-1) \right ].
\label{deltaLL}
    \end{equation}
We note here that the term $\left(1+\frac{P}{\rho}\right)$, which indicates the difference between $\Delta \rho$ and $\Delta n$, can be ignored in most parts of the star. It is approximately $1.2$ in the center and equal to $1$ at the surface. To calculate $d\ln T/dm$ we use Kramers' law for the opacity \begin{equation}
    k=k_0 \rho^{\nu}T^{-s}\phi,
\end{equation} 
where $k_0$ is the appropriate constant in the usual Kramers' opacity law suppressed by a factor $\phi$ that we introduce in order to account for the fact that the dark sector doesn't necessarily have the same unit charge as electromagnetism. One can now integrate
\begin{equation}
    \frac{dL}{dm}=\epsilon=\epsilon_0 \rho^{\beta},
    \end{equation}
   to find luminosity, and subsequently integrate Eq.~\eqref{eq:tempm} to determine the temperature of the star as function of the mass (or radius), which yields
\begin{align}
T(m)&=\left(T^{s+4}(0)-\chi(m)\right)^{1/(s+4)}~\text{with} \\
\chi(m)&=\frac{3k_0 \epsilon_0 (s+4)}{64 \pi^2 \alpha }\int\limits_0^m dm'\frac{\rho^{\nu}(m')}{r^4(m')}\int\limits_0^{m'} dm'' \rho^\beta (m'')\nonumber.
\end{align}
From Eq.~(\ref{epsilon_DP}) in the case of s-wave $\epsilon_0=f\langle \sigma_{\rm ann} v\rangle /m_{\chi}$. 
For the oscillation to set off we must have $W>0$, i.e., a  positive net work produced by all the layers of the star. Integrating over a period of the pulsation and by using Eqs.~(\ref{workdone}), (\ref{deltaE}), (\ref{DeltaLm}) and (\ref{deltaLL}), $W$ is positive if the following is also positive
\begin{align}\label{eq:instint}
A&=\int\limits_{0}^{M}dm {\left( \frac{\Delta n}{n}  \right)}^{2} \epsilon \left[ \epsilon_{\rho}+\epsilon_{T}(\Gamma_{3}-1)-(4+s)(\Gamma_{3}-1)+\kappa_{\rho}\right]-4\int\limits_{0}^{M}dm\, \epsilon  \left( \frac{\Delta n}{n}  \right) \left(\frac{\delta r}{r}\right)\nonumber\\
&\phantom{=}- \int\limits_{0}^{M}dm\; L\left( \frac{\Delta n}{n}  \right)  \left[(4+s)(\Gamma_{3}-1)-\kappa_{\rho}\right]\frac{\partial}{\partial m} \left(\frac{\Delta n}{n}\right)-4 \int\limits_{0}^{M}dm\; L\left( \frac{\Delta n}{n}  \right) \frac{\partial}{\partial m}\left(\frac{\delta r}{r}\right)\nonumber\\
&\phantom{=} -\left[\frac{\Delta n}{n} L (\Gamma_3-1)\frac{T}{dT/dm}\frac{\partial}{\partial m}\left(\frac{\Delta n}{n}\right)\right]\Bigg|_{\text{surface}}+\int\limits_{0}^{M}dm \left(\frac{\partial}{\partial m}\left(\frac{\Delta n}{n}\right)\right)^{2}L (\Gamma_3-1)\frac{T}{dT/dm}. 
\end{align}
We assume that $\Gamma_3 -1$ remains constant throughout the star, as it is expected to vary only slightly. In the last term of Eq.~(\ref{deltaLL}), we integrate by parts the temperature contribution. As anticipated, the star's temperature decreases slightly throughout the interior, making the inverse of the derivative a very large negative number. Thus, the primary contributions come from the last two terms of Eq.~(\ref{eq:instint}), which involve the temperature derivative. Between the two, the surface term gives numerically the dominant contribution, by three orders of magnitude, ensuring that $A$ is positive.

\begin{figure}[hbtp!]\label{fig:deltas}
   \centering
 \includegraphics[scale=0.7]{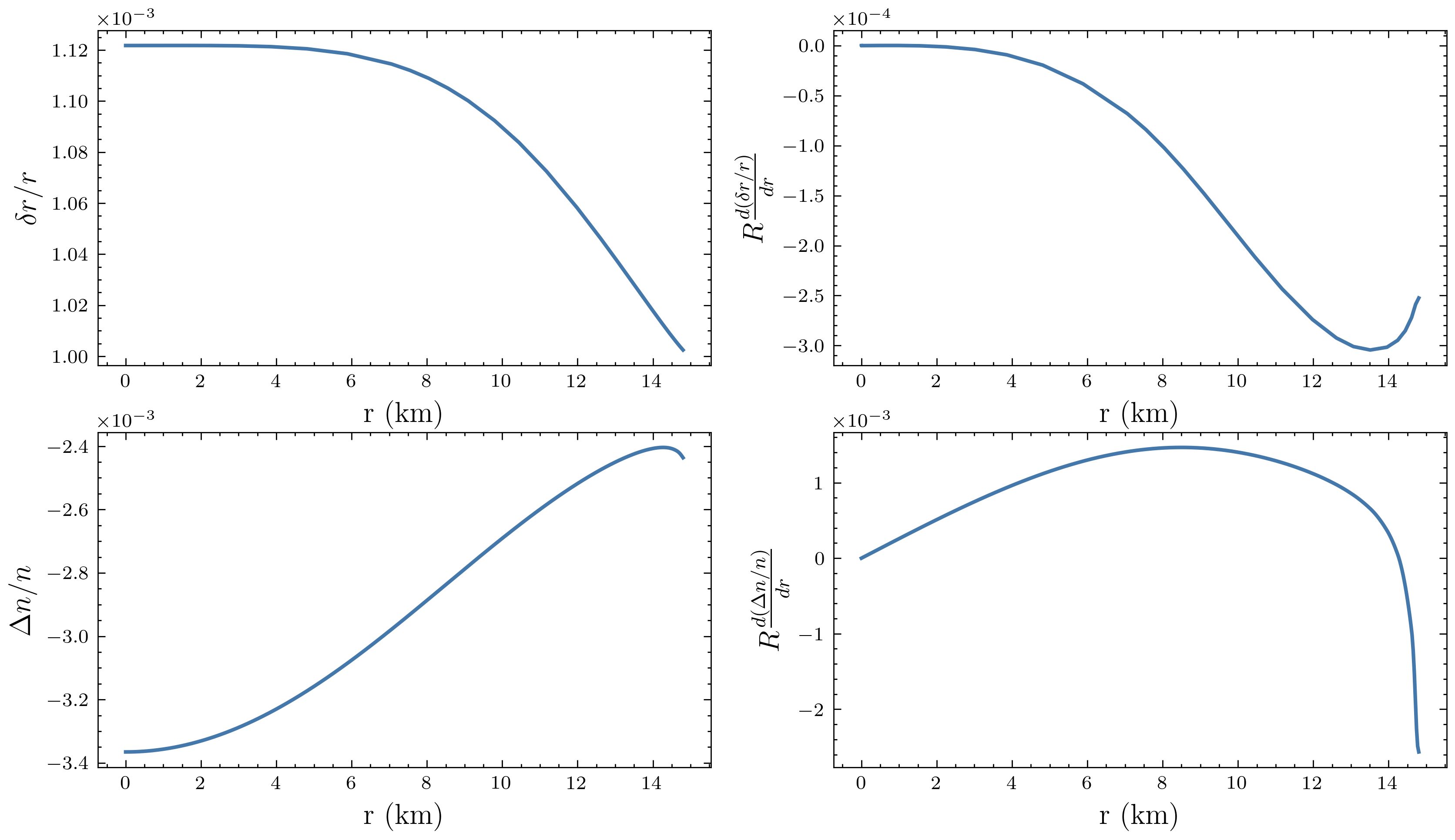}
 \caption{Relevant perturbations that affect the sign of $A$ (see text) in Eq.~(\ref{eq:instint}) as functions of $r$.}\label{fig:Instability}
\end{figure}

It is also interesting  to investigate  the contribution of the remaining terms of Eq.~(\ref{eq:instint}) if we completely ignore the last two temperature dependent terms. 
 The remaining terms give contributions of relatively similar size. This is reflected by the fact that the small quantities involved in each term, i.e., $\Delta n/n$, $\delta r/r$, $M \frac{d}{dm}(\Delta n/n)$ and $M \frac{d}{dm}(\delta r/r)$ are roughly at the same order of magnitude. Fig.~\ref{fig:Instability} depicts  the relevant quantities throughout the star (the third benchmark star from Table~\ref{tab:gaussian_fit}). Therefore the sign of the overall contribution depends on the delicate  coefficients that appear in each of the first four terms of Eq.~(\ref{eq:instint}). Fig.~\ref{fig:Instability} shows derivatives of perturbations with respect to $r$ instead of $m$ that appear in  Eq.~(\ref{eq:instint}). However this does not alter the conclusions. We can easily move from one derivative to the other recalling that $dm=4\pi r^2 \rho dr$. 
For s-wave annihilation $\epsilon_{\rho}=1,\;\epsilon_t=0$ and for Kramers  opacity we have $s=7/2,\; k_{\rho}=1$. Assuming $\Gamma_3-1=1/3$ which is the case for a non-interacting relativistic DM particle,  we estimate the bracket of the first term to be
\begin{equation}\label{kappas}
\epsilon_{\rho}-(4+s)(\Gamma_{3}-1)+k_{\rho}=-0.5.
\end{equation}
Given that the other terms seem to come with negative contributions one might think $A$ is negative. However, for the fundamental mode $\Delta n$ and $\delta r$ have the opposite sign, as shown in Fig.~\ref{fig:Instability}. This difference in phase simply means that compression $\delta r<0$, makes the density higher $\Delta n>0$.
The contribution of $\delta r/r$ comes with the largest coefficient, and in most stars, this is enough to overcome both the first and fourth negative terms. Additionally, some configurations can lead to a negative $\frac{d}{dr}(\Delta n/n)$ near the surface, making the third term positive as well (see Fig.~\ref{fig:Instability}). The energy produced for s-wave annihilation is given by $\epsilon=\epsilon_0 \rho$, where $\epsilon_0$ can be derived from \eqref{epsilon_DP}.
We evaluated $A$ for the third benchmark star using and $f=1\%$ and $\langle \sigma_{\rm ann}v\rangle=10^{-66}\;\text{cm}^{2}$.   Since $A$ has units of luminosity, we normalize it with respect to the average luminosity of the star $L_{m}$. We find that the first four terms give a positive $A/L_{m}\sim 10^{-5}$ while the 2 temperature terms an $A/L_{m}\sim 10^{5}$.  It is apparent that the temperature terms, particularly the surface term, dominate the other contributions. $W>0$ can still occur even without the temperature terms, so the $\epsilon$-mechanism is at play in these stars. Due to this net work produced after each period, radial oscillations can persist longer than other perturbations, making them easier to detect. It is worth mentioning that p-wave or d-wave annihilation enhances $A$ even further because the corresponding Eq.~(\ref{kappas}) increases and changes sign. Similar conclusions are drawn from the other three benchmark stars.

\section{Conclusions}\label{7}
In this paper we have investigated the possibility that asymmetric dark matter forms compact objects which can undergo radial oscillations. 
We assume that DM is made of degenerate fermions that can possess a Yukawa repulsive interaction.
We studied the radial oscillations within the full relativistic framework. If the DM conserved quantum number is violated at a high energy scale (e.g. the Planck scale), DM annihilation even with tiny cross sections could lead to an observable $\gamma$-ray spectrum which is well approximated by a Gaussian. This spectrum is shaped by the interplay between increasing Fermi energy as we move deeper in the star (thus increasing the annihilation rate and emitted energy) and the gravitational redshift which reduces the photon energy more in the star's core than at the surface. Annihilation cross sections suppressed by large energy scales which have negligible rates when applied to free particles, have in the contrary significant and observable rates in the dense environment of a compact object. Since the annihilation rate is modulated by the particle number density, radial oscillations  lead to a modulation of the produced luminosity with a frequency that of the excited radial mode, which lies for the studied parameters at the kHz range (or less).

There are several ways to induce radial oscillations. A close by encounter with another compact object can excite non-radial modes due to tidal effects. These modes will produce gravitational waves that eventually will damp the non-radial modes. However before that, non-radial modes can also trigger radial modes via nonlinear effects. The latter do not produce gravitational waves and therefore can last longer depending on the dissipation mechanisms present in the star. Another more interesting possibility is to have an internal mechanism that triggers the radial oscillations, like the case of certain variable stars. We found that under generic conditions, the DM annihilations throughout the star can put in work an $\epsilon$-mechanism. In that case radial oscillations do not attenuate with time and the interplay between the energy released via annihilation and the dissipation mechanisms only sets the amplitude of the oscillation. In ordinary stars  the $\epsilon$-mechanism is not effective on triggering oscillations because the energy released inside the star comes from nuclear fusion which is constrained in the core of the stars. Dissipation leads to oscillation damping in the outer shells. It is for this reason that pulsating stars with no large mass can achieve oscillations via another mechanism that involves changes in the opacity law.
$\epsilon$-mechanism can only be efficient in supermassive stars. In the case of the dark stars we studied, this is not the case. Because DM annihilation is not limited in the star's core but it is present throughout the star, $\epsilon$-mechanism can cause sustainable oscillations for stars out of the mass range where this mechanism is efficient for ordinary stars. 

Potential observation of variable stars with a Gaussian $\gamma$-ray spectrum, masses of the order of the mass of the sun and modulation frequencies in kHz, could be an indication towards this scenario. This potentially could not only give us a glimpse of the nature of DM but it can be a unique portal to indirectly study high energy scales beyond the Standard Model. 

There are several  directions that we leave for future work. One is to study non-radial oscillations in the same context. This is also relevant from the point of view of gravitational wave production. Another direction is to study the shear viscosity and thermal conductivity of these asymmetric stars in order to estimate precisely the potential amplitude of the radial oscillations.

\bibliographystyle{JHEP}
\bibliography{Draft}

\end{document}